\documentclass[aps,twocolumn,showpacs,superscriptaddress]{revtex4}

\usepackage{graphicx}
\usepackage{bm}
\usepackage{amsmath}
\usepackage{dcolumn}%

\newcommand{\eg}{\ensuremath{e_{g}}}
\newcommand{\tg}{\ensuremath{t_{2g}}}

\newcommand{\mb}{\ensuremath{\mu_{\text{B}}}}

\newcolumntype{/}{D{/}{/}{2,2}}  
\newcolumntype{.}{D{.}{.}{0}}  

\begin{document}

\title{Electronic structure and resonant inelastic x-ray scattering in
  Ca$_3$Ru$_2$O$_7$ }

\author{V.N. Antonov}

\affiliation{G. V. Kurdyumov Institute for Metal Physics of the
  N.A.S. of Ukraine, 36 Academician Vernadsky Boulevard, UA-03142
  Kyiv, Ukraine}

\affiliation{Max-Planck-Institute for Solid State Research,
  Heisenbergstrasse 1, 70569 Stuttgart, Germany}

\author{D.A. Kukusta}

\affiliation{G. V. Kurdyumov Institute for Metal Physics of the
  N.A.S. of Ukraine, 36 Academician Vernadsky Boulevard, UA-03142
  Kyiv, Ukraine}

\author{L.V. Bekenov}

\affiliation{G. V. Kurdyumov Institute for Metal Physics of the
  N.A.S. of Ukraine, 36 Academician Vernadsky Boulevard, UA-03142
  Kyiv, Ukraine}

\date{\today}

\begin{abstract}

  We have investigated the electronic structure of the transition
  metal oxide Ca$_3$Ru$_2$O$_7$ within density-functional theory using
  the generalized gradient approximation while considering strong
  Coulomb correlations in the framework of the fully relativistic
  spin-polarized Dirac linear muffin-tin orbital band-structure
  method. Ca$_3$Ru$_2$O$_7$ can be classified as a Mott insulator
  since it was expected to be metallic from band structure
  calculations. We found that the magnetic ground state of
  Ca$_3$Ru$_2$O$_7$ possesses an AFM-$b$ magnetic structure with the
  Ru spin moments ordered antiferromagnetically along the $b$ axis. We
  have investigated the resonant inelastic x-ray scattering (RIXS)
  spectra at the Ru and Ca $K$, $L_3$, and $M_3$ edges as well as at
  the O $K$ edge. The experimentally measured RIXS spectrum of
  Ca$_3$Ru$_2$O$_7$ at the Ru $L_3$ edge possesses a sharp feature
  $\le$2 eV corresponding to transitions within the Ru {\tg}
  levels. The excitation located from 2 eV to 4 eV is due to {\tg}
  $\rightarrow$ {\eg} transitions. The third wide structure situated
  at 4.5$-$11 eV appears due to transitions between the Ru
  4$d_{\rm{O}}$ states derived from the tails of oxygen 2$p$ states
  and {\eg} and {\tg} states. The RIXS spectra at the Ru $L_3$ and
  $M_3$ edges are very similar. However, the corresponding RIXS
  spectra at the Ca site quite differ from each other due to the
  significant difference in the widths of core-levels. The RIXS
  spectrum at the O $K$ edge consists of three major inelastic
  excitations. We found that the first two low energy features
  $\le$2.5 eV are due to interband transitions between occupied and
  empty O$_{\tg}$ states which appear due to the strong hybridization
  between oxygen 2$p$ and Ru {\tg} states in close vicinity to the
  Fermi level. The next major peak between 2.5 and 8 eV reflects
  interband transitions from occupied O 2$p$ to empty O$_{t_{2g}}$
  states. A wide energy interval between 4 and 11 eV is occupied by
  rather weak O$_{2p}$ $\rightarrow$ O$_{e_g}$ transitions.

\end{abstract}


\maketitle

\section{Introduction}

\label{sec:introd}

Ruddlesden-Popper (RP) type ruthenates A$_{n+1}$Ru$_n$O$_{3n+1}$ (A =
Sr or Ca), where $n$ is a number of Ru-O layers per unit cell, have
attracted much attention due to their fascinated physical
properties. For example, Sr$_2$RuO$_4$ ($n$ = 1) possesses a unique
$p$-wave superconductivity \cite{MHY+94,MRS01}, however, Ca$_2$RuO$_4$
is a typical Mott insulator \cite{CMS+97}. In the $n$ = $\infty$
family SrRuO$_3$ is a ferromagnetic metal \cite{CMS+97}, whereas
CaRuO$_3$ does not show any magnetic ordering
\cite{KAG+99}. Double-layered Sr$_3$Ru$_2$O$_7$ ($n$ = 2) has an
orthorhombic crystal structure (space group $Bbcb$, No. 68)
\cite{SJC+00}, where RuO$_6$ octahedra rotate around the $c$ axis. The
ionic radius of Ca$^{2+}$ is smaller than that of Sr$^{2+}$,
therefore, Ca$_3$Ru$_2$O$_7$ has a more distorted crystal structure,
namely, the orthorhombic symmetry with $Bb2_1m$ space group (No. 36)
with rotation and tilting of RuO$_6$ octahedra
\cite{CAM+00,YIM+05}. Sr$_3$Ru$_2$O$_7$ is a paramagnetic metal with
unusual large quasiparticle masses \cite{BGF+06}, it shows the
Fermi-liquid behavior, and ferromagnetic instability in the ground
state \cite{IMN+00}. On the other hand, Ca$_3$Ru$_2$O$_7$ is a
Mott-like antiferromagnetic (AFM) insulator, where ferromagnetic
bilayers stack antiparallel along the $c$ axis (N$\acute{\rm{e}}$el
temperature $T_N$ = 56 K) \cite{CMC+97,KHP+11}. This ruthenate
possesses the first-order metal-to-insulator transition at $T_{MI}$ =
48 K that distorts the RuO$_6$ octahedra and coincides with an upturn
in the out-of-plane resistivity \cite{OYI+04,YIM+05}. A spin rotation
from the $a$ axis to the $b$ axis (AFM-$b$) occurs at $T_{MI}$ that
appears to be mediated by an incommensurate spin state
\cite{YNI+04,DVF+20}. The system also shows complex magnetic field
dependence \cite{BMQ+08,SKH+19}. Ca$_3$Ru$_2$O$_7$ stimulated much
research effort due to its colossal magnetoresistance and anisotropic
magnetic behavior \cite{CMC+97,MCC+98,CBX+03,OYI+04,CLB+04}. This
bilayer compound also retains highly anisotropic electric resistivity
down to the lowest temperatures \cite{YNI+04}. Specific-heat and
photoemission studies \cite{PSB+98,CMC+99} indicate unusual behavior
for the density of states (DOSs) and bandwidth of Ca$_3$Ru$_2$O$_7$ at
low temperatures, reflecting strong electron correlation effects in
the 4$d$ band. The optical conductivity shows a small energy gap of 25
meV at low temperature \cite{LMY+07}. The Raman spectra measurements
by Liu {\it et al.} \cite{LYC+99} reveal a charge energy gap of
$\Delta_c$ $\sim$ 96 meV in Ca$_3$Ru$_2$O$_7$. The large gap ratio
$\Delta_c/k_B T_{MI}$ = 23 associated with this gap is indicative of
strong electronic correlations, perhaps pointing to a
Mott-Hubbard-type transition. It is believed that the wide spread of
physical phenomena in these ruthenates are due to a delicate balance
of competing interactions, such as, orbital degrees of freedom,
crystal structure distortion, electron correlation, and spin-orbit
coupling (SOC) \cite{MaSi97,MTS+01,FNT04,LAJ+08,HET+08,BKS+21}.

There are several band-structure calculations of Ca$_3$Ru$_2$O$_7$ based on
the density-functional theory (DFT) \cite{SiAu06,Liu11,JiKu18,MWC+20,PHC+20}.
Singh and Auluck provided the band structure calculations of Ca$_3$Ru$_2$O$_7$ and
found that the local spin-density approximation (LSDA) failed to reproduce the
insulating ground state \cite{SiAu06}. Liu reported an electronic structure
study on Ca$_3$Ru$_2$O$_7$ using first-principles calculations including the
SOC and on-site Coulomb interaction \cite{Liu11}. He found that the observed
insulating ground state of Ca$_3$Ru$_2$O$_7$ can be reproduced if the magnetic
moments are aligned along the $b$ axis.

In the present study, we focus on the electronic structure and
resonant inelastic x-ray scattering (RIXS) spectra of
Ca$_3$Ru$_2$O$_7$. The RIXS measurements at the Ru $L_3$ edge for the
Ca$_3$Ru$_2$O$_7$ ruthenate have been successfully performed recently
by Bertinshaw {\it et al.}  \cite{BKS+21}. They capture the in-plane
magnon across the entire Brillouin zone, giving a spin-wave gap of
$\sim$8 meV. At a higher energy loss, they observe $dd$-type
excitations between the {\tg} and {\eg} bands. The measurements of the
oxygen $K$ RIXS spectrum have been provided in Ref. \cite{AFH+20}.
The authors resolve two intra-{\tg} excitations in
Ca$_3$Ru$_2$O$_7$. The lowest lying excitation is interpreted as a
magnetic transverse mode with multiparticle character. They also
observe a very intensive peak at $\sim$3.3 eV in the O $K$ RIXS
spectrum.

In this paper, we report a detailed theoretical DFT study of the electronic
structure and RIXS spectra of the Ca$_3$Ru$_2$O$_7$ oxide at the Ru and Ca
$K$, $L_3$, and $M_3$ edges as well as at the O $K$ edge to investigate the
influence of the SOC and Coulomb correlations. The energy band structure of 
this RP ruthenate was calculated using the fully relativistic spin-polarized Dirac 
linear muffin-tin orbital band-structure method. We used both the generalized gradient
approximation (GGA) and the GGA+$U$ approach to assess the sensitivity of the
RIXS results to different treatment of the correlated electrons.

This paper is organized as follows. The crystal structures of Ca$_3$Ru$_2$O$_7$
and computational details are presented in Sec. II. Section III presents the
electronic and magnetic structures of this ruthenate. In Sec. IV, the
theoretical investigations of the RIXS spectra of Ca$_3$Ru$_2$O$_7$ at the Ru
and Ca $K$, $L_3$ edges and at the O $K$ edge are presented. The theoretical
results are compared with available experimental data. Finally, the results
are summarized in Sec. V.

\section{Computational details}
\label{sec:details}

\subsection{Resonant inelastic x-ray scattering.} 

In the direct RIXS process \cite{AVD+11}, the incoming photon with
energy $\hbar \omega_{\mathbf{k}}$, momentum $\hbar \mathbf{k}$, and
polarization $\bm{\epsilon}$ excites the solid from the ground state
$|{\rm g}\rangle$ with energy $E_{\rm g}$ to the intermediate state
$|{\rm I}\rangle$ with energy $E_{\rm I}$. During relaxation, an
outgoing photon with energy $\hbar \omega_{\mathbf{k}'}$, momentum
$\hbar \mathbf{k}'$, and polarization $\bm{\epsilon}'$ is emitted, and
the solid is in state $|f \rangle$ with energy $E_{\rm f}$. As a
result, an excitation with energy
$\hbar \omega = \hbar \omega_{\mathbf{k}} - \hbar
\omega_{\mathbf{k}'}$ and momentum $\hbar \mathbf{q}$ =
$\hbar \mathbf{k} - \hbar \mathbf{k}'$ is created.  Our implementation
of the code for the calculation of the RIXS intensity uses Dirac
four-component basis functions \cite{NKA+83} in the perturbative
approach \cite{ASG97}. RIXS is the second-order process, and its
intensity is given by

\begin{eqnarray}
I(\omega, \mathbf{k}, \mathbf{k}', \bm{\epsilon}, \bm{\epsilon}')
&\propto&\sum_{\rm f}\left| \sum_{\rm I}{\langle{\rm
    f}|\hat{H}'_{\mathbf{k}'\bm{\epsilon}'}|{\rm I}\rangle \langle{\rm
    I}|\hat{H}'_{\mathbf{k}\bm{\epsilon}}|{\rm g}\rangle\over
  E_{\rm g}-E_{\rm I}} \right|^2 \nonumber \\ && \times
\delta(E_{\rm f}-E_{\rm g}-\hbar\omega),
\label{I1}
\end{eqnarray}
where the RIXS perturbation operator in the dipole approximation is
given by the lattice sum $\hat{H}'_{\mathbf{k}\bm{\epsilon}}=
\sum_\mathbf{R}\hat{\bm{\alpha}}\bm{\epsilon} \exp(-{\rm
  i}\mathbf{k}\mathbf{R})$, where $\alpha$ are the Dirac matrices. The
sum over the intermediate states $|{\rm I}\rangle$ includes the
contributions from different spin-split core states at the given
absorption edge. The matrix elements of the RIXS process in the frame
of the fully relativistic Dirac LMTO method were presented in our
previous publication \cite{AKB22a}.

\subsection{X-ray magnetic circular dichroism} 

Magneto-optical (MO) effects refer to various changes in the polarization
state of light upon interaction with materials possessing a net magnetic
moment, including rotation of the plane of linearly polarized light (Faraday,
Kerr rotation), and the complementary differential absorption of left and
right circularly polarized light (circular dichroism). In the near visible
spectral range these effects result from excitation of electrons in the
conduction band. Near x-ray absorption edges, or resonances, MO effects can be
enhanced by transitions from well-defined atomic core levels to transition
symmetry selected valence states.

Within the one-particle approximation, the absorption coefficient
$\mu^{\lambda}_j (\omega)$ for incident x-ray polarization $\lambda$ and
photon energy $\hbar \omega$ can be determined as the probability of
electronic transitions from initial core states with the total angular
momentum $j$ to final unoccupied Bloch states

\begin{eqnarray}
\mu_j^{\lambda} (\omega) &=& \sum_{m_j} \sum_{n \bf k} | \langle \Psi_{n \bf k} |
\Pi _{\lambda} | \Psi_{jm_j} \rangle |^2 \delta (E _{n \bf k} - E_{jm_j} -
\hbar \omega ) \nonumber \\
&&\times \theta (E _{n \bf k} - E_{F} ) \, ,
\label{mu}
\end{eqnarray}
where $\Psi _{jm_j}$ and $E _{jm_j}$ are the wave function and the energy of a
core state with the projection of the total angular momentum $m_j$;
$\Psi_{n\bf k}$ and $E _{n \bf k}$ are the wave function and the energy of a
valence state in the $n$-th band with the wave vector {\bf k}; and $E_{F}$ is
the Fermi energy.

Here, $\Pi _{\lambda}$ is the electron-photon interaction
operator in the dipole approximation
\begin{equation}
\Pi _{\lambda} = -e \mbox{\boldmath$\alpha $} \bf {a_{\lambda}},
\label{Pi}
\end{equation}
where $\bm{\alpha}$ are the Dirac matrices and $\bf {a_{\lambda}}$ is the
$\lambda$ polarization unit vector of the photon vector potential,
with $a_{\pm} = 1/\sqrt{2} (1, \pm i, 0)$,
$a_{\parallel}=(0,0,1)$. Here, $+$ and $-$ denote, respectively, left
and right circular photon polarizations with respect to the
magnetization direction in the solid. Then, x-ray magnetic circular
and linear dichroisms are given by $\mu_{+}-\mu_{-}$ and
$\mu_{\parallel}-(\mu_{+}+\mu_{-})/2$, respectively.  More detailed
expressions of the matrix elements in the electric dipole
approximation may be found in
Refs.~\cite{GET+94,book:AHY04,AHS+04}.  The matrix elements due
to magnetic dipole and electric quadrupole corrections are presented
in Ref.~\cite{AHS+04}.

\subsection{Crystal structure.} 

\begin{figure}[tbp!]
\begin{center}
\includegraphics[width=1.\columnwidth]{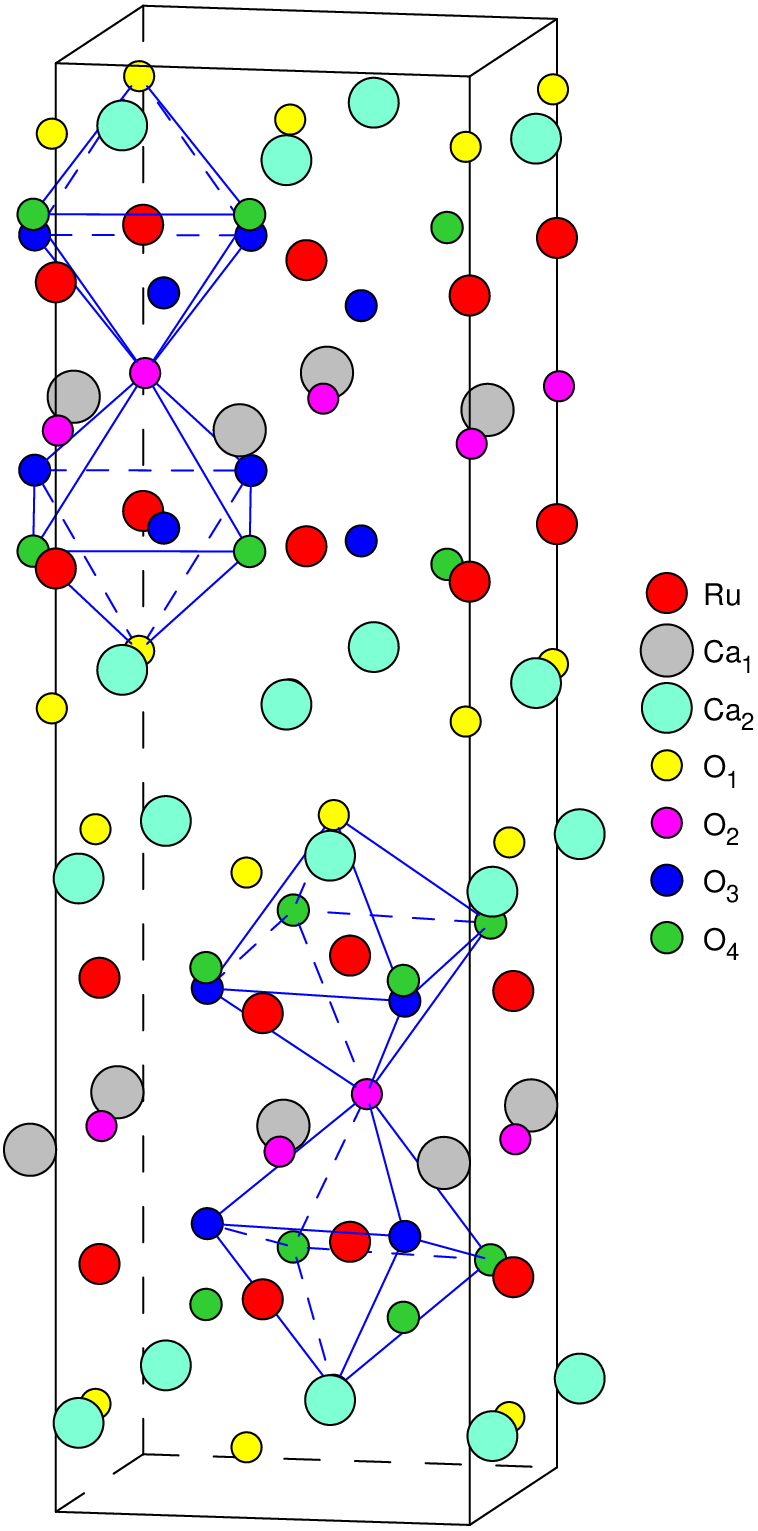}
\end{center}
\caption{\label{struc_CRO}(Color online) (upper panel) The crystal structure
  of Ca$_3$Ru$_2$O$_7$ (space group $Bb2_1m$, No. 36) \cite{YIM+05} with
  Ru$^{3+}$ ions in octahedral oxygen coordination. }
\end{figure}

Ca$_3$Ru$_2$O$_7$ has the orthorhombic symmetry with $Bb2_1m$ space group
(No. 36) with the rotation and tilting of RuO$_6$ octahedra \cite{CAM+00,YIM+05}
(see Table \ref{T_struc_CRO} and Fig. \ref{struc_CRO}). The structure of
Ca$_3$Ru$_2$O$_7$ consists of a RuO$_6$ double layer and CaO layer per unit
cell. The Ru$^{4+}$ cations are surrounded by oxygen octahedra with the
interatomic distances Ru-O$_1$, Ru-O$_2$, Ru-O$_3$, and Ru-O$_4$ of 1.9872,
1.9873, 2.0033, and 1.9949 \AA\,, respectively. The inter-atomic
distances between Ru and Ca$_1$ and Ca$_2$ ions are equal to 3.1263 and 3.0355
\AA\,, respectively.

\begin{table}[tbp!]
  \caption {The atomic as well as Wyckoff positions (WP) of
    Ca$_3$Ru$_2$O$_7$ with $Bb2_1/m$ lattice symmetry (lattice constants $a$ =
    5.3677 \AA, $b$ = 5.5356 \AA\, and $c$ = 19.5212 \AA) at temperature of 8
    K \cite{YIM+05}. }
\label{T_struc_CRO}
\begin{center}
\begin{tabular}{|c|c|c|c|c|}
\hline
 Atom &  WP & $x$      & $y$    & $z$     \\
\hline
 Ca$_1$ & 4$a$  & 0.7382     & 0.1949   &  0   \\
 Ca$_2$ & 8$b$  & 0.2443     & 0.3036   & 0.3112    \\
 Ru     & 8$b$  & 0.2533     & 0.7512   & 0.4011  \\
 O$_1$  & 8$b$  & 0.8188     & 0.2309   &  0.6987    \\
 O$_2$  & 4$a$  & 0.3370     & 0.7702   &  0.5    \\
 O$_3$  & 8$b$  & 0.4474     & 0.9500   &  0.0819    \\
 O$_4$  & 8$b$  & 0.9481     & 0.0458   &  0.1162    \\
\hline
\end{tabular}
\end{center}
\end{table}

The neutron diffraction measurements show that the lattice constants $a$ and
$c$ of Ca$_3$Ru$_2$O$_7$ shorten on cooling from room temperature (RT), while
$b$ elongates \cite{YIM+05}. At the first-order transition temperature of 48 K
the lattice parameters jump, though, the space group symmetry $Bb2_1m$ is
not affected throughout the temperature range measured \cite{YIM+05}. The $c$
constant shortens by about 0.1\%, while both $a$ and $b$ lengthen by about
0.07\% at $T_{MI}$.

The lengths of Ru-O$_1$ and Ru-O$_3$ bonds increase with decreasing
temperature. On the other hand, those of Ru-O$_2$ and Ru-O$_4$
decrease with decreasing temperature. The angles between the RuO$_6$
octahedra are almost temperature independent above $T_N$ and little
affected by the first-order phase transition. However, some
characteristic temperature dependences can be observed $\le$40 K
\cite{YIM+05}. The angle of Ru-O$_1$-Ca$_2$ decreases with decreasing
temperature. On the other hand, the angle of Ru-O$_2$-Ru increases
with decreasing temperature \cite{YIM+05}.

\subsection{Calculation details}

The details of the computational method are described in our previous papers
\cite{AJY+06,AHY+07b,AYJ10,AKB22a} and here we only mention several
aspects. The band structure calculations were performed using the fully
relativistic linear muffin-tin orbital (LMTO) method
\cite{And75,book:AHY04}. This implementation of the LMTO method uses
four-component basis functions constructed by solving the Dirac equation
inside an atomic sphere \cite{NKA+83}. The exchange-correlation functional of
the GGA type was used in the version of
Perdew, Burke, and Ernzerhof \cite{PBE96}. The Brillouin zone integration was
performed using the improved tetrahedron method \cite{BJA94}. The basis
consisted of Ca and Ru $s$, $p$, $d$, and $f$, and O $s$, $p$, and $d$ LMTOs.

To consider the electron-electron correlation effects, we use
in this work the relativistic generalization of the rotationally
invariant version of the LSDA+$U$ method \cite{YAF03} which takes into
account that, in the presence of spin-orbit coupling, the occupation
matrix of localized electrons becomes non-diagonal in spin
indexes. Hubbard $U$ was considered an external parameter and
varied from 0.8 to 3.8 eV for the 4$d$ Ru site. The values of
exchange Hund coupling used in our calculations were obtained from
constrained LSDA calculations \cite{DBZ+84,PEE98} and equal to
$J_H$=0.8 eV for Ru. Thus, the parameter $U_{eff}=U-J_H$, which
roughly determines the splitting between the lower and upper Hubbard
bands, varied between 0 and 3.0 eV for the 4$d$ metal site. We
adjusted the value of $U$ to achieve the best agreement with the
experiment.

The x-ray absorption spectroscopy (XAS), x-ray magnetic circular
dichroism (XMCD), and RIXS spectra were calculated taking into account
the exchange splitting of core levels.
The finite lifetime of a core hole was accounted for by folding the spectra
with a Lorentzian. The widths of core levels $\Gamma$ for Ca, Ru, and O were
taken from Ref. \cite{CaPa01}. The finite experimental resolution of the
spectrometer was accounted for by a Gaussian of 0.6 eV (the $s$ coefficient of
the Gaussian function).

\section{Electronic and magnetic structures}
\label{sec:bands}

\begin{figure}[tbp!]
\begin{center}
\includegraphics[width=0.99\columnwidth]{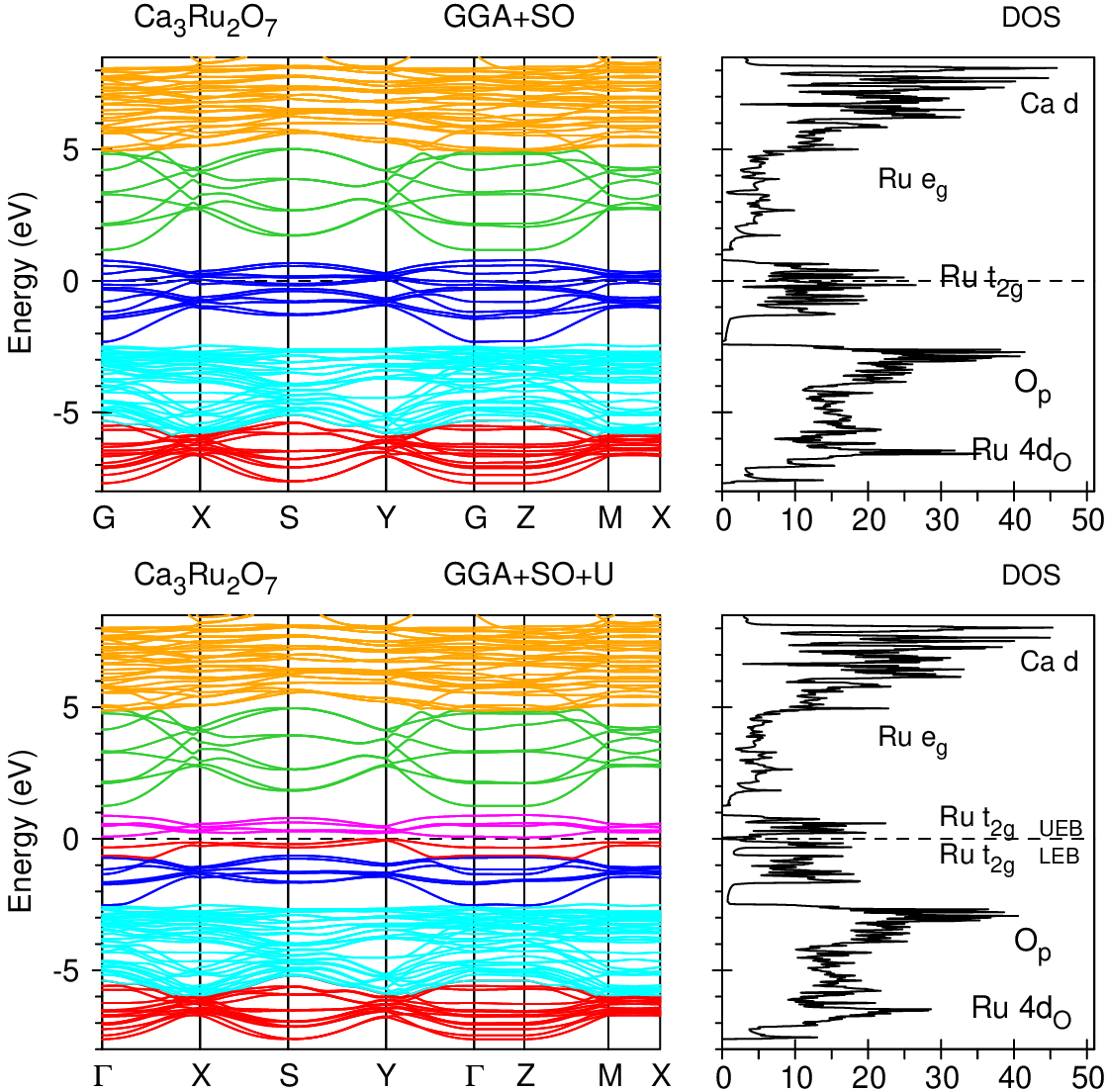}
\end{center}
\caption{\label{BND_CRO}(Color online) The energy band structure and total density of states 
  (DOS) [in states/(cell eV)] of Ca$_3$Ru$_2$O$_7$ calculated in the GGA+SO
  approximation (the upper panel) and GGA+SO+$U$ approach (the lower panel). }
\end{figure}

Figure \ref{BND_CRO} presents the energy band structure of Ca$_3$Ru$_2$O$_7$
calculated in the fully relativistic Dirac GGA+SO approximation (the upper
panel) and considering Coulomb correlations in the GGA+SO+$U$
approach (the lower panel). The GGA+SO approximation produces a metallic
ground state in Ca$_3$Ru$_2$O$_7$ in contradiction with the experimental data
which show that Ca$_3$Ru$_2$O$_7$ is a Mott-like AFM insulator below $T_{MI}$
= 48 K \cite{CMC+97,OYI+04,YIM+05,KHP+11}. To produce the correct ground state one
has to take into account strong Coulomb correlations in Ca$_3$Ru$_2$O$_7$. The
GGA+SO+$U$ approach shifts the occupied and empty {\tg} bands downward
and upward, respectively, by $U_{eff}$/2 producing a direct energy gap of
0.215 eV and an indirect one of 0.053 eV for $U_{eff}$ = 1.3 eV. We found that
this value of Hubbard $U$ applied for the Ru site produces the best agreement
between the calculated and experimentally measured RIXS spectra at the Ru
$L_3$ edge in Ca$_3$Ru$_2$O$_7$. The energy gap is increased with increasing
Hubbard $U$.

\begin{figure}[tbp!]
\begin{center}
\includegraphics[width=0.89\columnwidth]{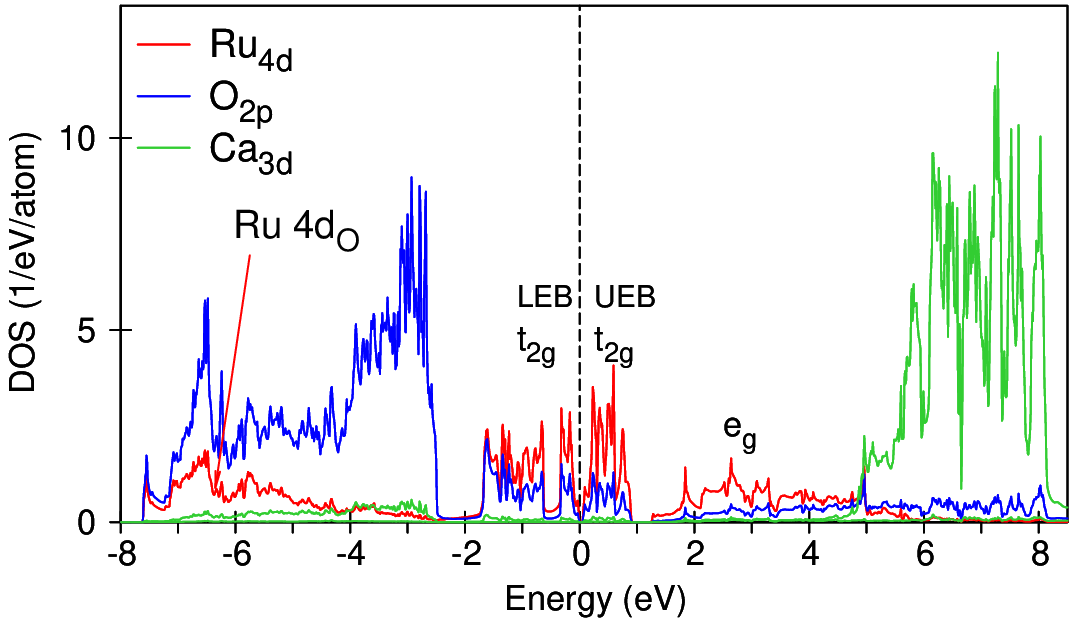}
\end{center}
\caption{\label{PDOS_CRO}(Color online) The partial density of states (DOS)
  in Ca$_3$Ru$_2$O$_7$ calculated in the GGA+SO+$U$ approach with
  $U_{eff}$ = 1.3 eV.}
\end{figure}

Figure \ref{PDOS_CRO} presents the partial density of states (DOS) in Ca$_3$Ru$_2$O$_7$
calculated in the GGA+SO+$U$ approach. Four electrons occupy the
{\tg}-type low energy band (LEB) manifold in the energy interval from $-$1.7
eV to $-$0.6 eV and between $-$0.35 eV and $E_F$ in Ca$_3$Ru$_2$O$_7$. The
empty {\tg} states [the upper energy band (UEB)] occupy the energy range from
0.1 eV to 0.9 eV. The {\eg}-type states of Ru are distributed far above the
Fermi level from 1.3 eV to 5.0 eV. The occupation number of 4$d$ electrons in
the Ru atomic sphere is equal to 5.66, which is much larger than the expected
value of four {\tg} electrons in Ru$^{4+}$. The excessive charge is
provided by the tails of O 2$p$ states inside the Ru atomic spheres. These
charge transfer 4$d_{\rm{O}}$ states, which are located from $-$7.6 to
approximately $-$5.0 eV, play an essential role for the RIXS spectrum at the
Ru $L_{2,3}$ edges (see Section IV). The Ca 3$d$ states are mostly empty and
located from 5.0 to 8.2 eV above the Fermi level. The oxygen 2$p$ states are
strongly hybridized with the Ru 4$d$ states due to small interatomic O-Ru
distances and delocalized nature of O 2$p$ states. They occupy the energy
interval between $-$7.6 eV and $-$2.5 eV. The small peaks in the vicinity 
of the Fermi level are due to the hybridization between O 2$p$ and Ru
{\tg} LEB and UEB.

\begin{figure}[tbp!]
\begin{center}
\includegraphics[width=0.9\columnwidth]{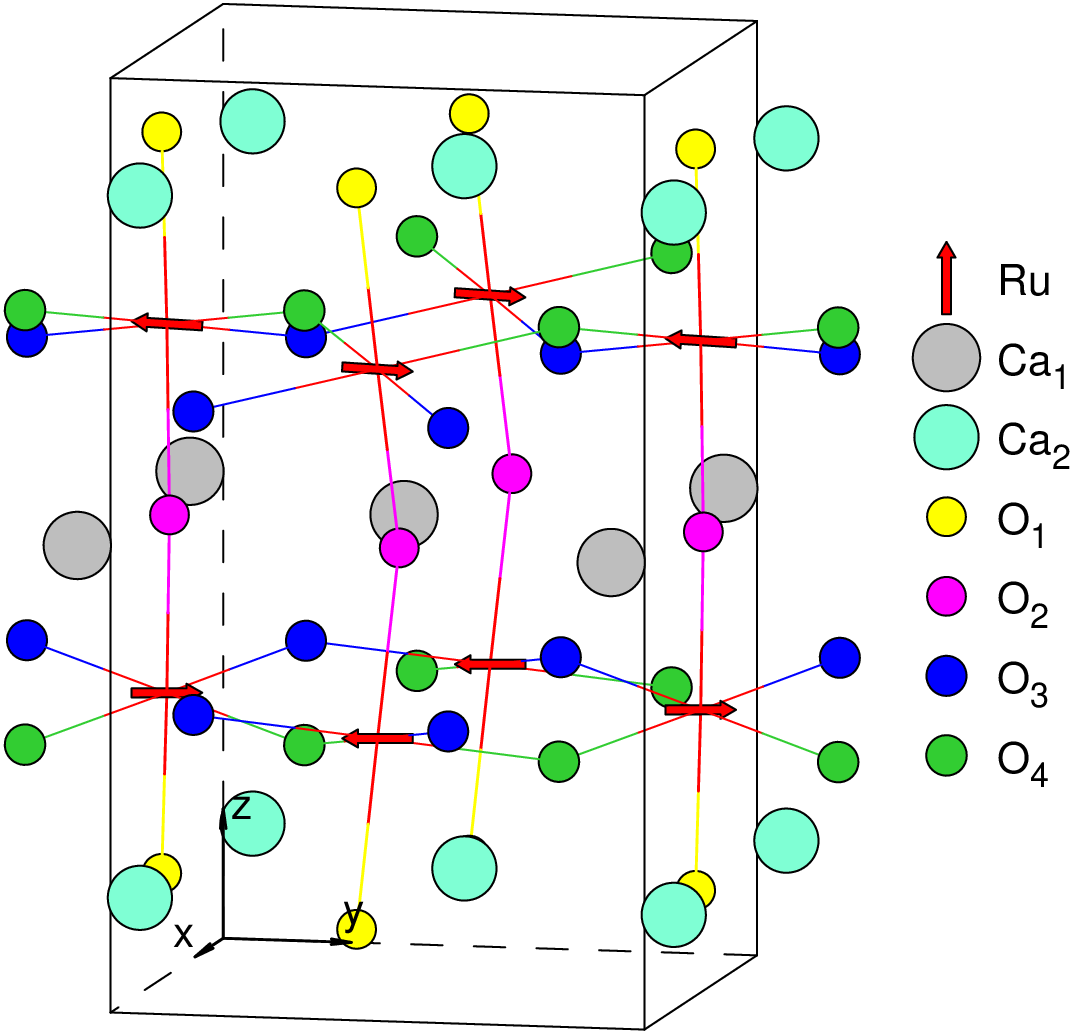}
\end{center}
\caption{\label{mag_CRO}(Color online) A half of the Ca$_3$Ru$_2$O$_7$ unit cell
  with AFM-ordered Ru moments parallel to the $b$ axis (the AFM-$b$ order). }
\end{figure}

We found that in the ground state Ca$_3$Ru$_2$O$_7$ possesses an AFM-$b$ magnetic
structure with the Ru spin moments ordered antiferromagnetically along the $b$
axis (see Fig. \ref{mag_CRO}). It is in consistence with experimental data
\cite{YNI+04,DVF+20} and previous band structure calculations \cite{Liu11}.

\begin{table}[tbp!]
  \caption{\label{mom_CRO} The theoretically calculated in GGA+SO+$U$ ($U_{eff}$ = 1.3 eV) spin $M_s$, orbital
    $M_l$, and total $M_{tot}$ magnetic moments (in {\mb}) in Ca$_3$Ru$_2$O$_7$. }
\begin{center}
\begin{tabular}{|c|c|c|c|}
\hline
Atom & $M_s$  & $M_l$ & $M_{total}$    \\
\hline
 Ca$_1$& 0.0002  & 0.0007   & 0.0009  \\
 Ca$_2$& 0.0165  & 0.0014   & 0.0179  \\
 Ru    & 1.4783  & 0.1370   & 1.6153  \\
 O$_1$ & 0.0625  & 0.0072   & 0.0697  \\
 O$_2$ & 0.0120  &$-$0.0065 & 0.0055  \\
 O$_3$ & 0.1008  &   0.0097 & 0.1105  \\
 O$_4$ & 0.1141  &   0.0096 & 0.1237  \\
\hline
\end{tabular}
\end{center}
\end{table}

Table \ref{mom_CRO} presents the theoretically calculated spin $M_s$,
orbital $M_l$, and total $M_{tot}$ magnetic moments in
Ca$_3$Ru$_2$O$_7$. The Ru spin and orbital moments are parallel in
Ca$_3$Ru$_2$O$_7$ in accordance with Hund's third rule. The spin and
orbital parts of the magnetic moments for the Ca ions in
Ca$_3$Ru$_2$O$_7$ are found to be very small. The spin and orbital
magnetic moments at the oxygen sites are also relatively small with
the largest moments at the O$_4$ site and the smallest ones at the O$_2$
site.

\section{XAS, XMCD, \lowercase{and} RIXS \lowercase{spectra}}
\label{sec:rixsL}

\subsection{R\lowercase{u} $L_3$ \lowercase{and} $M_3$ XAS, XMCD,
  \lowercase{and} RIXS \lowercase{spectra}}

\begin{figure}[tbp!]
\begin{center}
\includegraphics[width=0.9\columnwidth]{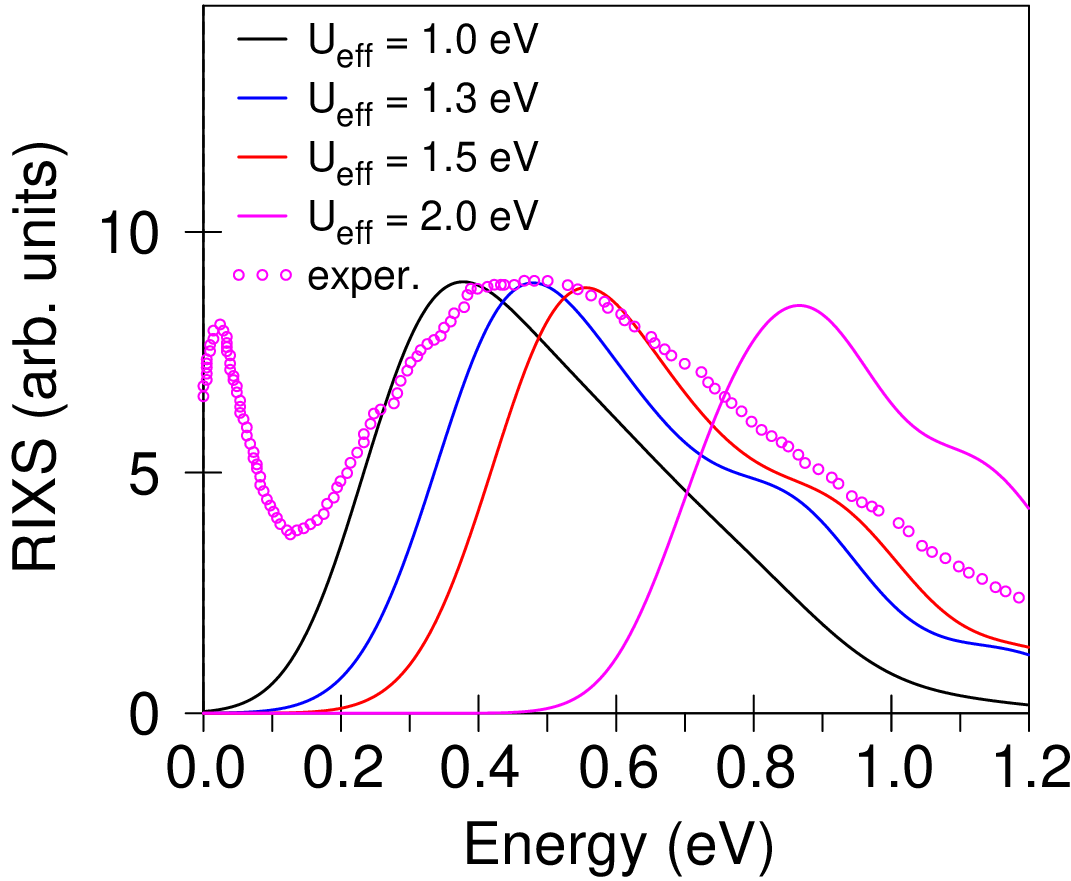}
\end{center}
\caption{\label{rixs_U_CRO}(Color online) The experimental RIXS
  spectrum measured by Bertinshaw {\it et al.} \cite{BKS+21} (open
  magenta circles) at the Ru $L_3$ edge of Ca$_3$Ru$_2$O$_7$ compared
  with the theoretical spectra calculated for $\tg \rightarrow \tg$
  transitions with different $U_{eff}$.  }
\end{figure}

The experimental RIXS spectrum at the Ru $L_3$ edge was measured by
Bertinshaw {\it et al.} \cite{BKS+21} in the energy range up to 6.5
eV. In addition to an elastic peak centered close to zero energy loss,
the spectrum consists of a peak at $\sim$0.5 eV with a high energy
shoulder at 1.3 eV and two peaks at 3 and 5.8 eV. We found that the
two peak structure situated $\le$2 eV corresponds to intra-{\tg}
excitations. These peaks are very sensitive to the value of the energy
gap in Ca$_3$Ru$_2$O$_7$ and the relative position of the {\tg} LEB
and UEB bands (Fig. \ref{BND_CRO}). Figure \ref{rixs_U_CRO} shows the
experimental RIXS spectrum measured by Bertinshaw {\it et al.}
\cite{BKS+21} (open magenta circles) in comparison with the
theoretical spectra calculated for $\tg \to \tg$ transitions with
different values of $U_{eff}$. The best agreement was found for the
GGA+SO+$U$ approach with $U_{eff}$ = 1.3 eV. The calculations with
larger $U_{eff}$ shift the RIXS spectrum towards higher energies.

\begin{figure}[tbp!]
\begin{center}
\includegraphics[width=0.9\columnwidth]{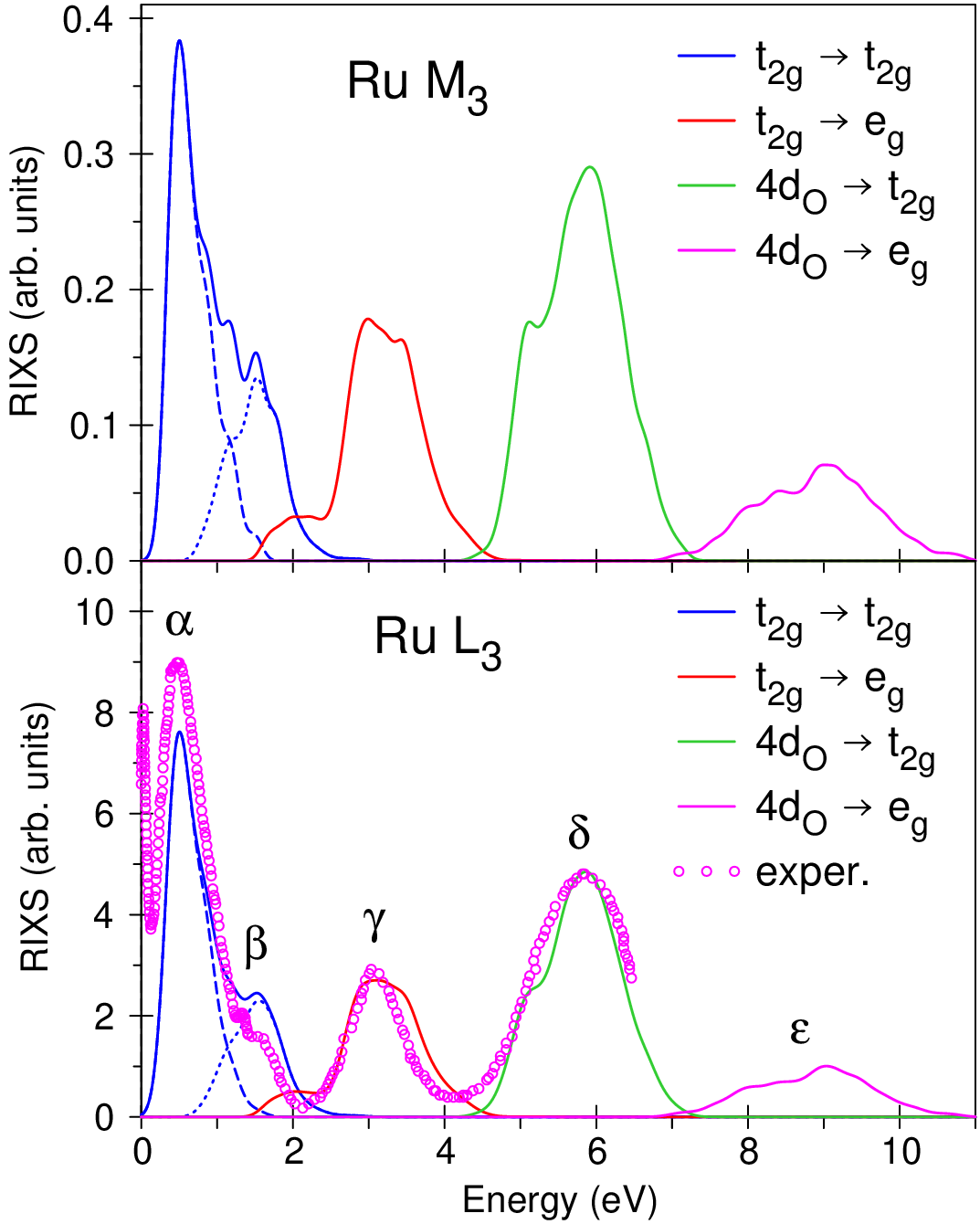}
\end{center}
\caption{\label{rixs_Ru_LM_CRO}(Color online) The experimental RIXS
  spectrum (open magenta circles) measured by Bertinshaw {\it et al.}
  \cite{BKS+21} at the Ru $L_3$ edge of Ca$_3$Ru$_2$O$_7$ compared
  with the theoretically calculated partial contributions to the RIXS
  spectrum from different interband transitions (the lower panel). The
  upper panel presents the theoretically calculated partial
  contributions to the RIXS spectrum from different interband
  transitions at the $M_3$ edge. The calculations were carried out in
  the GGA+SO+$U$ approach with $U_{eff}$= 1.3 eV. }
\end{figure}

Figure \ref{rixs_Ru_LM_CRO} (the lower panel) presents the
experimental RIXS spectrum (open magenta circles) at the Ru $L_3$
edges for Ca$_3$Ru$_2$O$_7$ \cite{BKS+21} in a wide energy interval in
comparison with the theoretically calculated partial contributions
from different interband transitions in the GGA+SO+$U$ approach. As we
mentioned above, the two peaks situated $\le$2 eV correspond to
intra-{\tg} excitations (dashed and dotted blue curves in
Fig. \ref{rixs_Ru_LM_CRO}). We divided the occupied part of the {\tg}
energy band into two groups: four bands in close vicinity to the Fermi
level (blue curves in Fig. \ref{BND_CRO}) and the other 12 bands
situated at lower energy (red curves in Fig. \ref{BND_CRO}); and
calculated the interband transitions from these two groups to empty
{\tg} states (magenta curves in Fig. \ref{BND_CRO}) separately. We
found that the small number of occupied bands in close vicinity to the
Fermi level produce a larger RIXS signal (peak $\alpha$) than the
large number of low energy bands (peak $\beta$) due to corresponding
matrix elements. The peak $\gamma$ located at $\sim$3 eV (the red
curve in the lower panel of Fig. \ref{rixs_Ru_LM_CRO}) was found to be
due to $\tg \rightarrow \eg$ transitions. The strong fine structure
$\delta$ at $\sim$5.8 eV (the green curve) is due to 4$d_{\rm{O}}$
$\rightarrow$ {\tg} transitions. The structure which extends from 7 up
to 11 eV (the magenta curve) presents 4$d_{\rm{O}}$ $\rightarrow$
{\eg} transitions. The theoretical calculations are in good agreement
with the experimental data.

Figure \ref{rixs_Ru_LM_CRO} (the upper panel) shows the
theoretically calculated RIXS spectrum at the Ru $M_3$ edge presented
as partial contributions from different interband transitions. The
shape of the spectrum is very similar to the corresponding spectrum at
the Ru $L_3$ edge, however, with approximately one order of magnitude
smaller intensity. 

\begin{figure}[tbp!]
\begin{center}
\includegraphics[width=0.9\columnwidth]{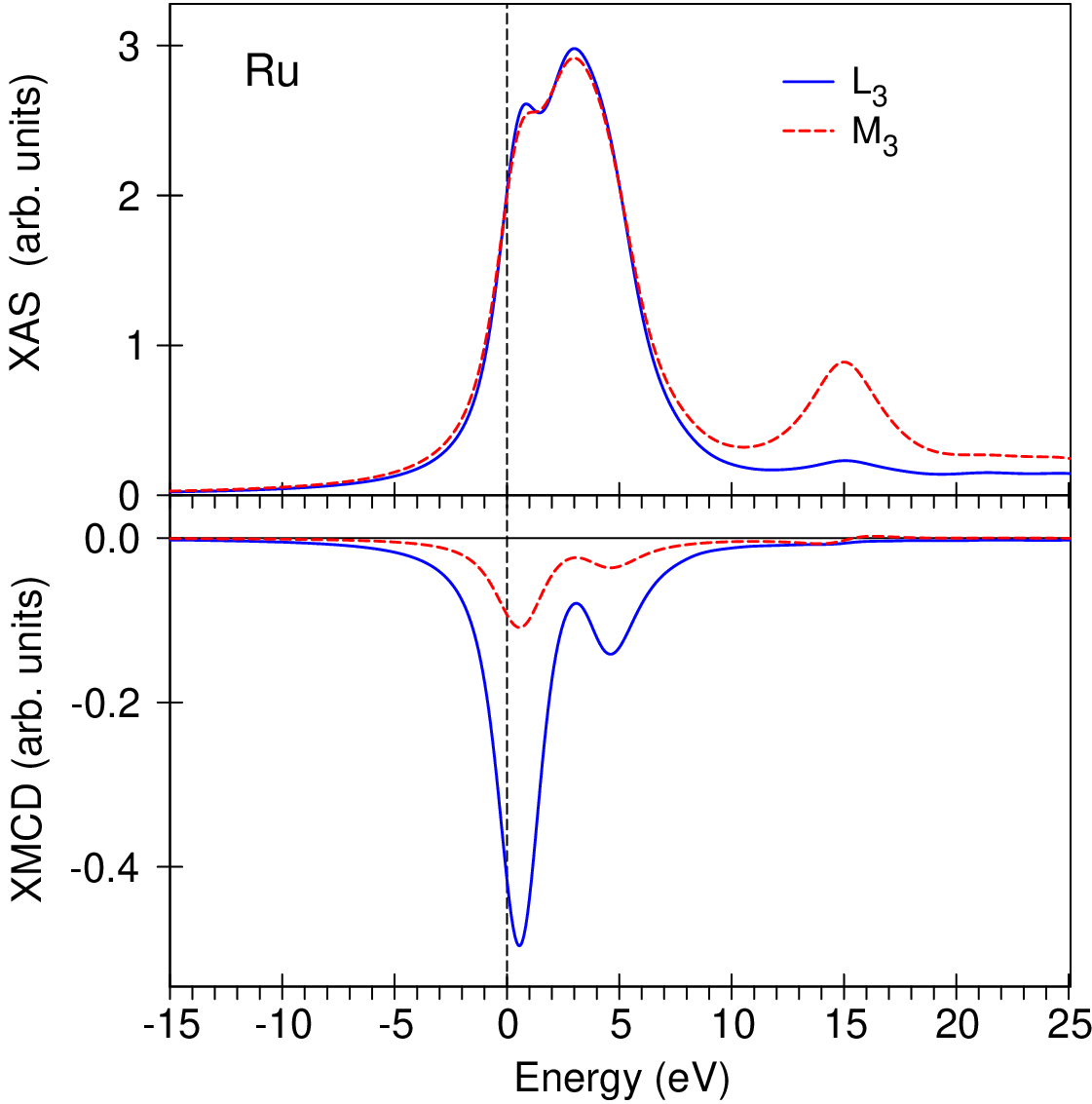}
\end{center}
\caption{\label{xmcd_Ru_LM_CRO}(Color online) The theoretically
  calculated XAS (the upper panel) and XMCD (the lower panel) spectra
  at the Ru $L_3$ and $M_3$ edges of Ca$_3$Ru$_2$O$_7$ calculated in
  the GGA+SO+$U$ approach with $U_{eff}$= 1.3 eV. }
\end{figure}

Figure \ref{xmcd_Ru_LM_CRO} shows the theoretically calculated XAS (the upper
panel) and XMCD (the lower panel) spectra at the Ru $L_3$ and $M_3$ edges of
Ca$_3$Ru$_2$O$_7$ calculated in the GGA+SO+$U$ approach. The widths of
the 2$p_{3/2}$ and 3$p_{3/2}$ core levels are close ($\Gamma_{2p_{3/2}}$ =
1.87 eV and $\Gamma_{3p_{3/2}}$ = 2.2 eV \cite{CaPa01}). Therefore, both the
Ru $L_3$ and $M_3$ XAS spectra possess very similar structures with a major peak
at $\sim$3 eV, a low energy shoulder at 0.7 eV, and an additional high energy
peak at 15 eV. The last fine structure is due to the hybridization of Ru
4$d$ states with the oxygen 2$p$ states which have a corresponding peak at 15
eV above the Fermi level (not shown). Two low energy fine structures in the Ru
$L_3$ ($M_3$) XAS spectrum $\le$5 eV are due to transitions from the core
2$p_{3/2}$ (3$p_{3/2}$) level to empty {\tg} (the low energy shoulder) and
{\eg} (the major peak) states. The transitions to {\eg} states have larger
intensity in comparison with the corresponding transitions into empty
{\tg} states. The XMCD spectra also have a two-peak structure (the lower panel
of Fig. \ref{xmcd_Ru_LM_CRO}) but these spectra are dominated by
transitions into empty {\tg} states with a small contribution from empty
{\eg} orbitals at higher energy. The XMCD spectrum at the $M_3$ edge is much
smaller than at the $L_3$ edge.

\subsection{C\lowercase{a} $L_3$ \lowercase{and} $M_3$ XAS, XMCD,
  \lowercase{and} RIXS \lowercase{spectra}}

\begin{figure}[tbp!]
\begin{center}
\includegraphics[width=0.9\columnwidth]{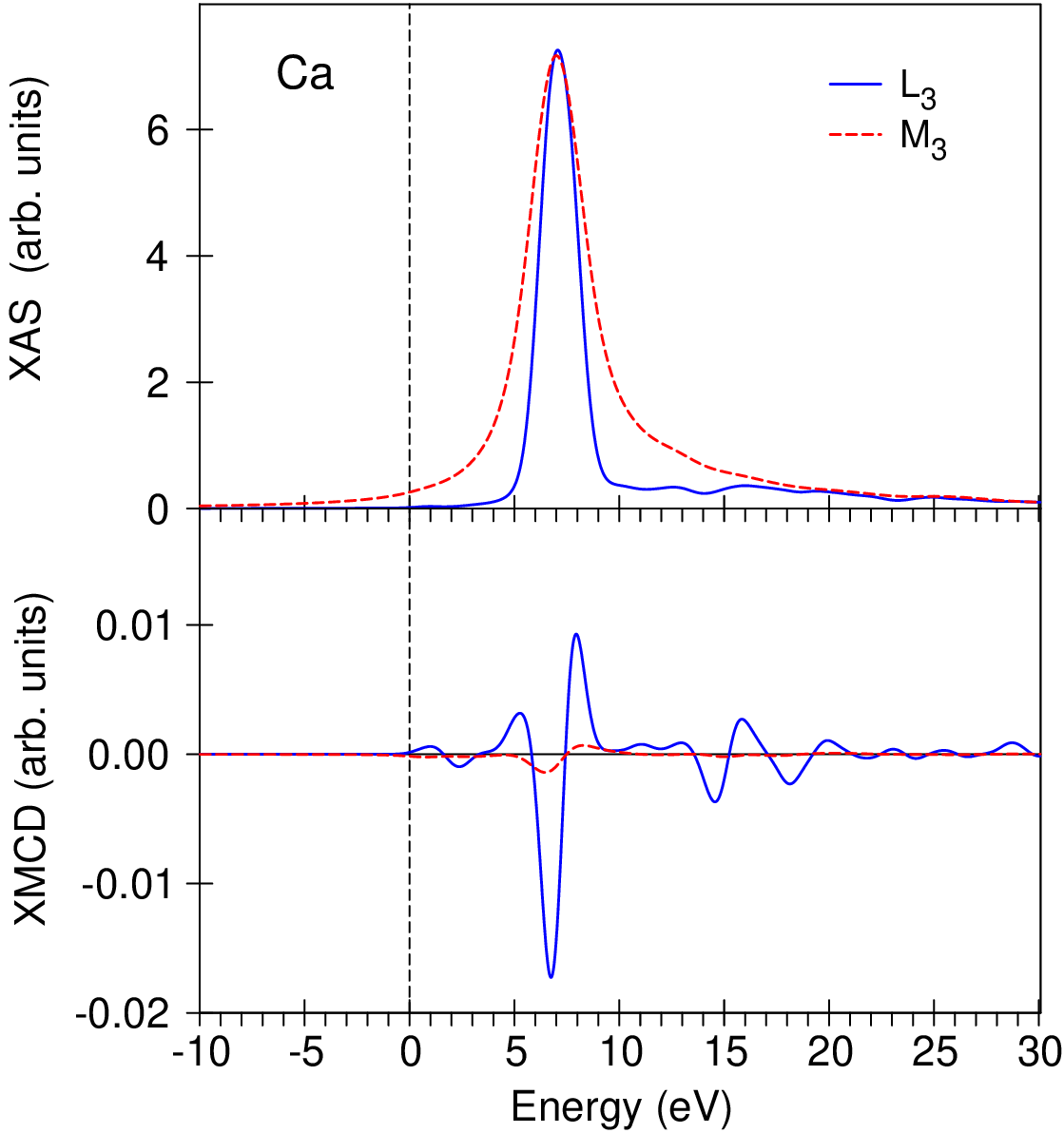}
\end{center}
\caption{\label{xmcd_Ca_LM_CRO}(Color online) The theoretically
  calculated XAS (the upper panel) and XMCD (the lower panel) spectra
  at the Ca $L_3$ and $M_3$ edges of Ca$_3$Ru$_2$O$_7$ calculated in
  the GGA+SO+$U$ approach with $U_{eff}$= 1.3 eV. }
\end{figure}

Figure \ref{xmcd_Ca_LM_CRO} shows the theoretically calculated XAS (the upper
panel) and XMCD (the lower panel) spectra at the Ca $L_3$ and $M_3$ edges of
Ca$_3$Ru$_2$O$_7$. The Ca 3$d$ partial DOS is situated far above the Fermi
level (see Fig. \ref{PDOS_CRO}). Therefore, both the Ca $L_3$ and $M_3$ XAS
spectra possess one-peak structures at $\sim$7 eV above the Fermi level. The
Ca $M_3$ XAS spectrum is much wider than the corresponding $L_3$ spectrum due
to a significantly larger width of the 3$p_{3/2}$ core level in comparison with
the 2$p_{3/2}$ one \cite{CaPa01}. The Ca $L_3$ XMCD spectrum has a complicated
shape with several positive and negative peaks. It is more than one order of
magnitude smaller than the Ru $L_3$ XMCD spectrum (the lower panel of
Fig. \ref{xmcd_Ru_LM_CRO}) due to smaller SOC, spin and orbital magnetic
moments at the Ca site in comparison with the Ru site (Table
\ref{mom_CRO}). The dichroism at the Ca $M_3$ edge is very small even
compared to the Ca $L_3$ XMCD spectrum.

\begin{figure}[tbp!]
\begin{center}
\includegraphics[width=0.9\columnwidth]{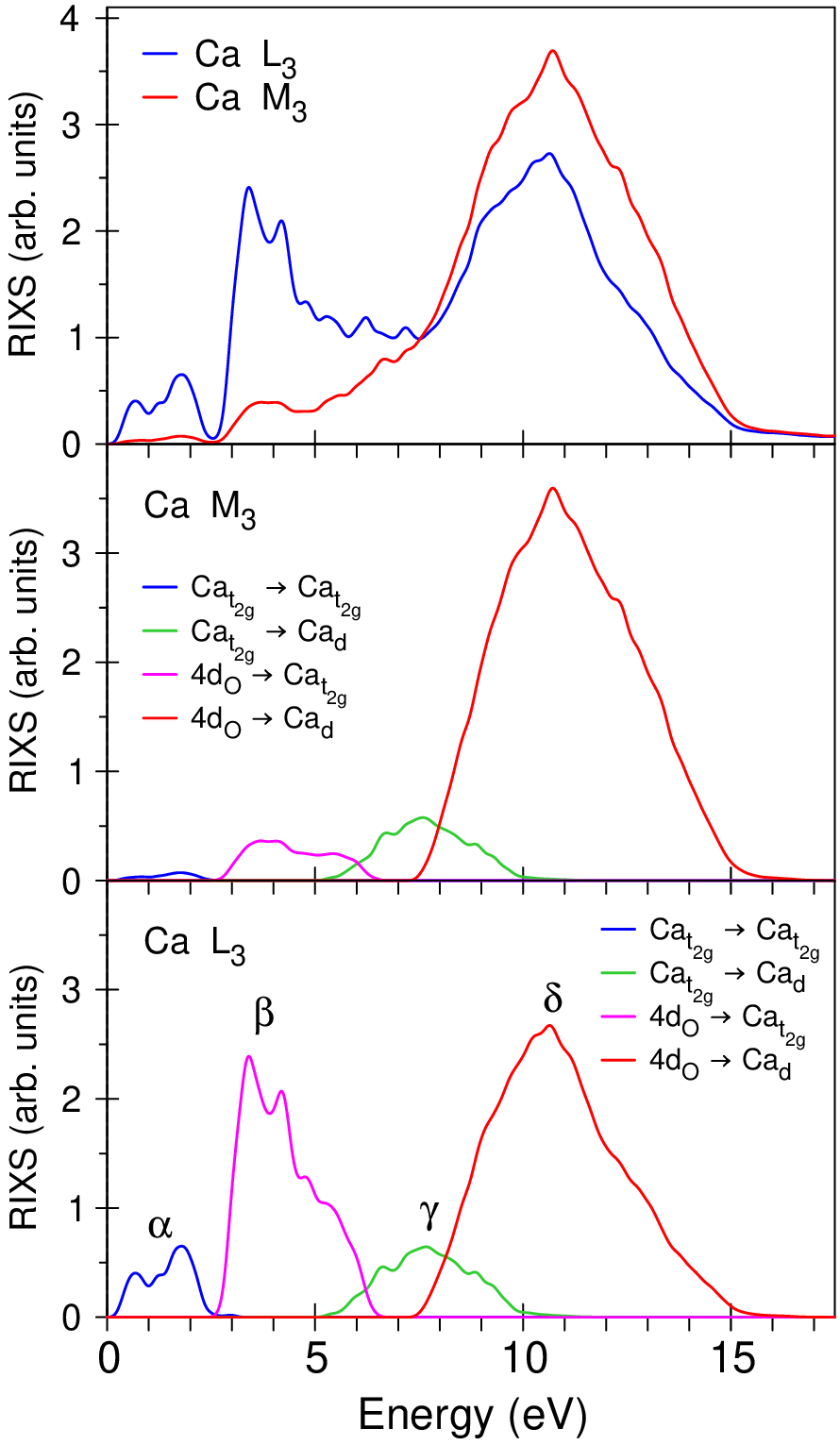}
\end{center}
\caption{\label{rixs_Ca_LM_CRO}(Color online) The theoretically calculated
  RIXS spectra at the Ca $L_3$ and $M_3$ edges in Ca$_3$Ru$_2$O$_7$ (the upper
  panel) calculated in the GGA+SO+$U$ approach ($U_{eff}$ = 1.3 eV) and
  partial contributions to the RIXS spectra from different interband
  transitions at the Ca $M_3$ (the middle panel) and Ca $L_3$ edges (the
  lower panel). }
\end{figure}

Figure \ref{rixs_Ca_LM_CRO} show the theoretically calculated RIXS spectra at
the Ca $L_3$ and $M_3$ edges in Ca$_3$Ru$_2$O$_7$ (the upper panel) and
partial contributions to the RIXS spectra from different interband transitions
at the Ca $M_3$ (the middle panel) and Ca $L_3$ edges (the lower
panel). The low energy two-peak fine structure $\alpha$ $\le$2.5 eV (the blue
curve) was found to be due to Ca$_{\tg} \rightarrow$ Ca$_{\tg}$
transitions. More precisely, between Ca 3$d$ occupied and empty states, which
are located in the energy region of Ru $\tg$ states from $-$1.7 eV to $E_F$
and from 0.1 to 0.9 eV, respectively (Fig. \ref{PDOS_CRO}). The next fine
structure $\beta$ located between 2.5 and 6.5 eV arises from interband
transitions between 4$d_O$ states (which are derived from the tails of oxygen 2$p$
states inside the Ca atomic spheres) and Ca$_{\tg}$ states. The peak
$\gamma$ is due to interband transitions from Ca$_{\tg}$ states to the Ca
empty 4$d$ states. The large high energy peak $\delta$ is due to
$4d_O \rightarrow$ Ca$_d$ transitions.

\begin{figure}[tbp!]
\begin{center}
\includegraphics[width=0.9\columnwidth]{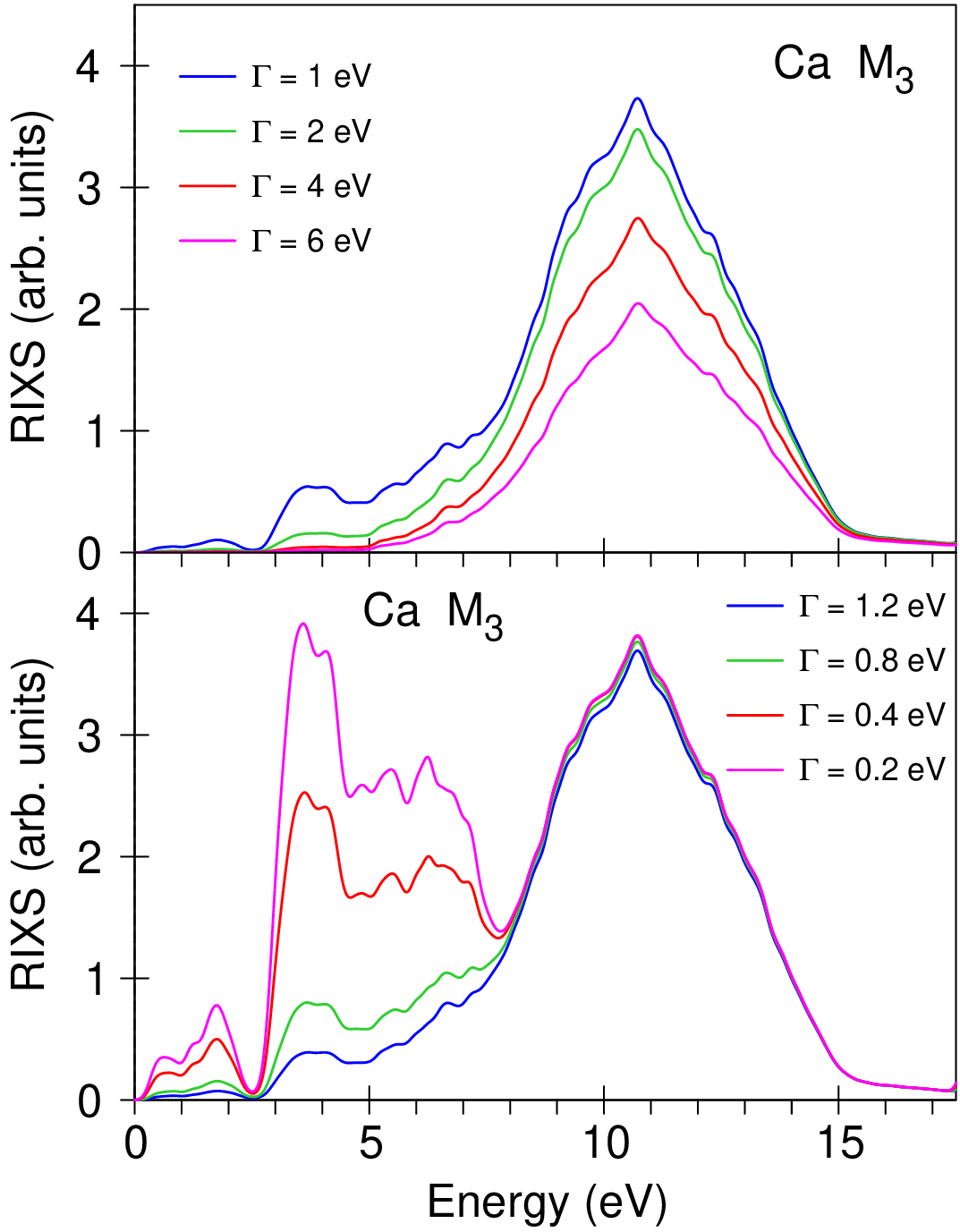}
\end{center}
\caption{\label{rixs_Ca_L3_gamma}(Color online) The theoretically
  calculated RIXS spectra at the Ca $M_3$ edge in Ca$_3$Ru$_2$O$_7$
  obtained for different parameters $\Gamma$. }
\end{figure}

The RIXS spectra at the $L_3$ and $M_3$ edges have to possess similar
shapes because they reflect interband transitions between $d_{5/2}$ band
states and the 2$p_{3/2}$ and 3$p_{3/2}$ core levels, respectively. This is
true indeed in the case of the Ru $L_3$ and $M_3$ RIXS spectra (see
Fig. \ref{rixs_Ru_LM_CRO}). However, the $L_3$ and $M_3$ RIXS spectra at the
Ca site differ from each other (the upper panel of
Fig. \ref{rixs_Ca_LM_CRO}). There is strong suppression of the low energy
peaks $\alpha$ and $\beta$ in the Ca $M_3$ RIXS spectrum in comparison with the Ca
$L_3$ spectrum. Such a puzzle can be explained by the significant difference in
the widths of corresponding core-levels ($\Gamma_{3p_{3/2}}$ = 1.2 eV and
$\Gamma_{2p_{3/2}}$ = 0.21 eV \cite{CaPa01}). Figure \ref{rixs_Ca_L3_gamma}
shows the theoretically calculated RIXS spectra at the Ca $M_3$ edge in
Ca$_3$Ru$_2$O$_7$ obtained for different core-level parameters $\Gamma$. Under
decreasing the parameter $\Gamma$ from 1.2 eV to 0.2 eV the intensity of the
low energy peaks $\alpha$ and $\beta$ is increased and the shape of the Ca
$M_3$ RIXS spectrum is transformed to the $L_3$ one. The high energy peak
$\delta$ does not change its shape. On the other hand, when we changed the parameter
$\Gamma$ to a greater extent from 1 to 6 eV we found that the intensity of the
peak $\delta$ monotonously decreased.

It should be mentioned that the x-ray absorption and RIXS processes
have different nature. The matrix elements of the RIXS process are
more complicated than the x-ray absorption ones
\cite{ASG97,book:AHY04,AKB22a}. As a result, the XAS spectra possess
linear dependence on the core level width: the larger $\Gamma_{core}$
the wider the XAS spectrum without a shape change (see the upper panel
of Fig. \ref{xmcd_Ca_LM_CRO}). The RIXS spectra show nonlinear
dependence on $\Gamma_{core}$, which strongly affects the spectrum
shape.

Experimental measurements of the XAS, XMCD, and RIXS spectra at
the Ca $L_3$, $M_3$, and Ru $M_3$ edges are highly desirable.

\subsection{O, C\lowercase{a}, \lowercase{and} R\lowercase{u} $K$ RIXS
\lowercase{spectra}}

The RIXS spectra at the O $K$ edge in Ca$_3$Ru$_2$O$_7$ were measured
by Arx {\it et al.} \cite{AFH+20} up to 5 eV. The O $K$ RIXS spectrum
consists of a peak centered at zero energy loss, which comprises the
elastic line and other low-energy features such as phonons, magnons,
etc., two inelastic excitations at 0.5 eV, 1.8 eV, and a major peak at
$\sim$3.3. We found that the first two low energy features are due to
interband transitions between occupied and empty O$_{\tg}$ states,
which appear as a result of the strong hybridization between oxygen
2$p$ states with Ru {\tg} LEB and UEB in the vicinity of the Fermi
level (see Fig. \ref{PDOS_CRO}). Therefore, the oxygen $K$ RIXS
spectroscopy can be used for the estimation of the energy band gap and
positions of Ru 4$d$ Hubbard bands. The major peak at 3.3 eV reflects
interband transitions from occupied O 2$p$ states to the empty oxygen
states which originate from the hybridization with Ru {\tg} states. We
found that the theory reproduces well the shape and energy position of
the two low energy features. The energy position of the major peak at
3.3 eV is also reproduced well by theory, although our calculations
give a wider major peak in comparison with the experiment. The
interband transitions O$_{2p} \rightarrow$ O$_{\eg}$ are relatively
small and occupy a wide energy interval from 4.1 to 11 eV.

\begin{figure}[tbp!]
\begin{center}
\includegraphics[width=0.9\columnwidth]{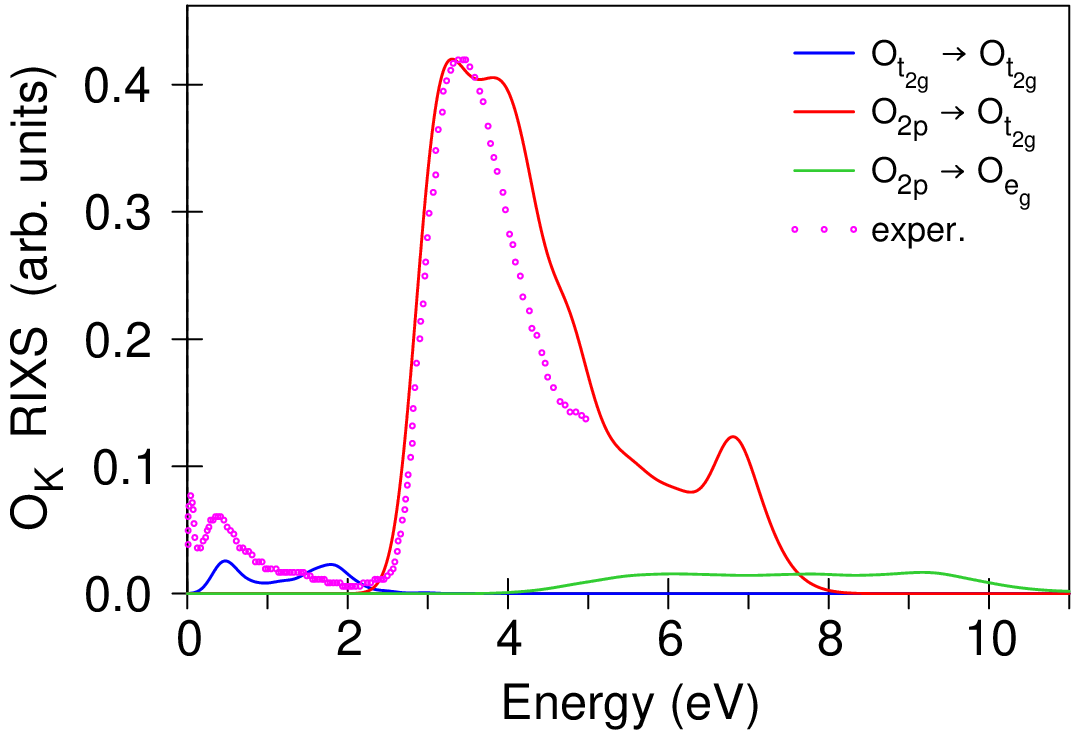}
\end{center}
\caption{\label{rixs_O_K_CRO}(Color online) The theoretically
  calculated RIXS spectrum at the O $K$ edge in Ca$_3$Ru$_2$O$_7$
  obtained in the GGA+SO+$U$ approach ($U_{eff}$ = 1.3 eV) and
  partial contributions to the RIXS spectrum from different interband
  transitions in comparison with the experiment \cite{AFH+20}. }
\end{figure}

\begin{figure}[tbp!]
\begin{center}
\includegraphics[width=0.9\columnwidth]{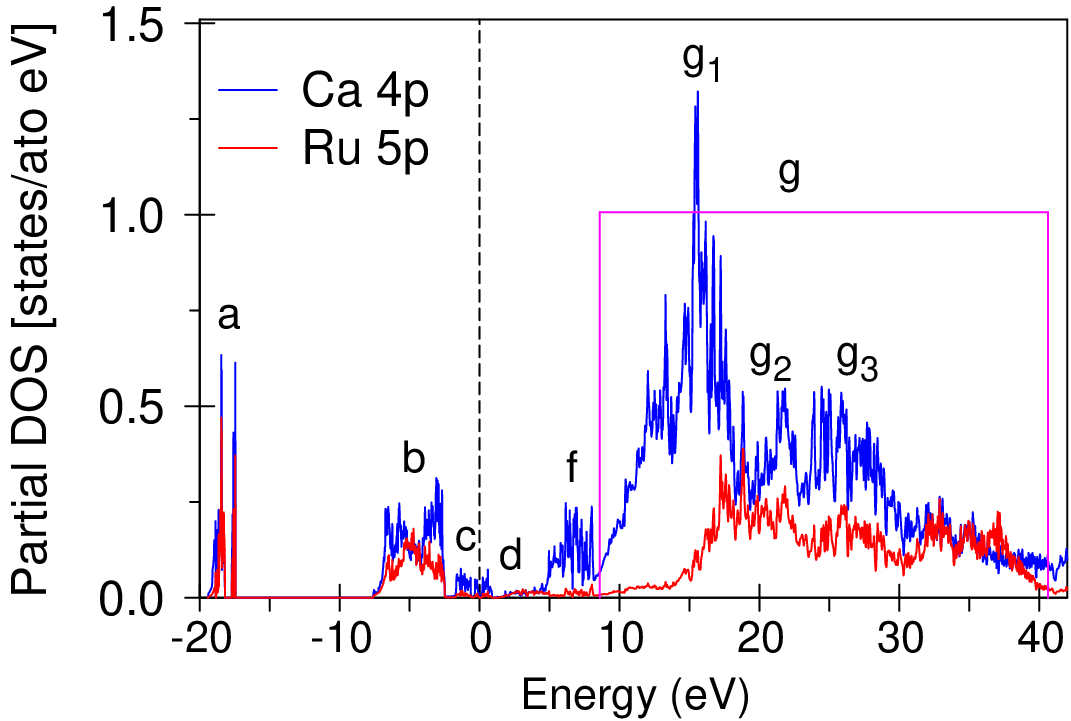}
\end{center}
\caption{\label{PDOS Ca_Ru_p}(Color online) The partial Ca 4$p$ and Ru 5$p$
  DOS [in states/(atom eV)] in Ca$_3$Ru$_2$O$_7$ calculated
  in the GGA+SO+$U$ ($U_{eff}$= 1.3 eV) approach. }
\end{figure}

Let us consider now the XAS, XMCD, and RIXS spectra at the Ru and Ca
$K$ edges.  For that we first present the partial Ca 4$p$ and Ru 5$p$
DOS in Fig. \ref{PDOS Ca_Ru_p} in a wide energy interval from $-$20 to
40 eV. We distinguish several groups of the bands. The group $a$
derives from the hybridizations of the $p$ states with corresponding Ca
4$s$ and Ru 5$s$ states. The group $b$ is due to the hybridization with
oxygen 2$p$ states. The groups $c$ and $d$ are from the hybridization
with Ru {\tg} LEB and UEB, respectively. The group $f$ comes from the
hybridization with Ru {\eg} states. At last, the very wide structure $g$,
which we divide into several subgroups, is the Ca 4$p$ and Ru 5$p$
bands themselves.

\begin{figure}[tbp!]
\begin{center}
\includegraphics[width=0.9\columnwidth]{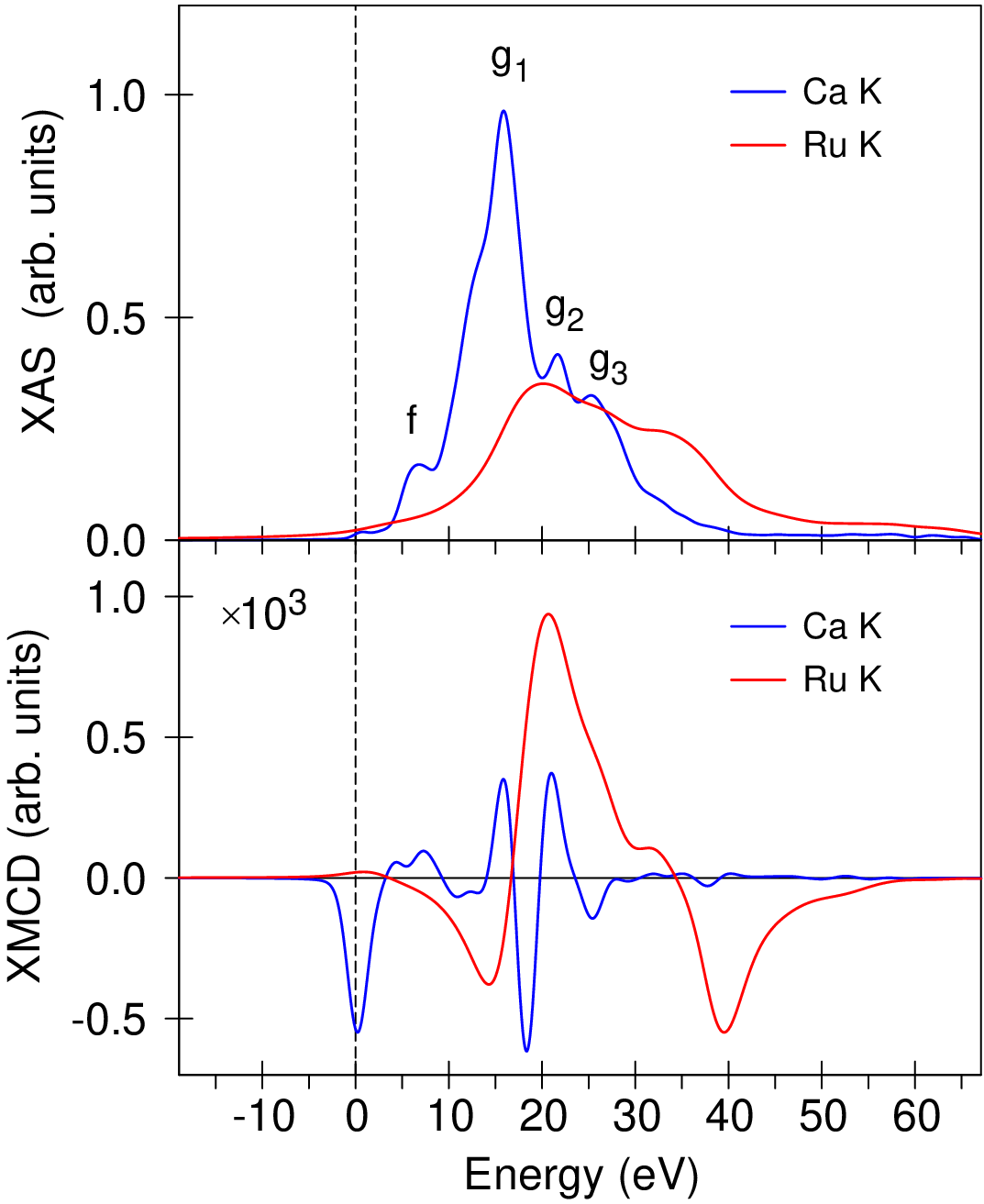}
\end{center}
\caption{\label{XAS_Ca_K_CRO}(Color online) The theoretically
  calculated XAS (the upper panel) and XMCD spectra (the lower
  panel) at the Ca and Ru $K$ edges in Ca$_3$Ru$_2$O$_7$ calculated in
  the GGA+SO+$U$ approach ($U_{eff}$ = 1.3 eV). }
\end{figure}

Figure \ref{XAS_Ca_K_CRO} presents the theoretically calculated XAS (the upper
panel) and XMCD (the lower panel) at the Ca and Ru $K$ edges. The four fine
structures $f$, $g_1$, $g_2$, and $g_3$ in the Ca $K$ XAS spectrum can be
easily distinguished as transitions from the Ca 1$s$ core level into
corresponding groups of the Ca 4$p$ states with the same labels. The width of
the Ru 1$s$ core level is much larger than the Ca one ($\Gamma^{\rm{Ru}}_{1s}$
= 5.33 eV and $\Gamma^{\rm{Ca}}_{1s}$ = 0.77 eV \cite{CaPa01}), therefore, the Ru
$K$ XAS spectrum possesses less structured features. The XMCD spectra at both
Ca and Ru $K$ edges are at least three order of magnitude smaller than the
corresponding XAS spectra.

\begin{figure}[tbp!]
\begin{center}
\includegraphics[width=0.9\columnwidth]{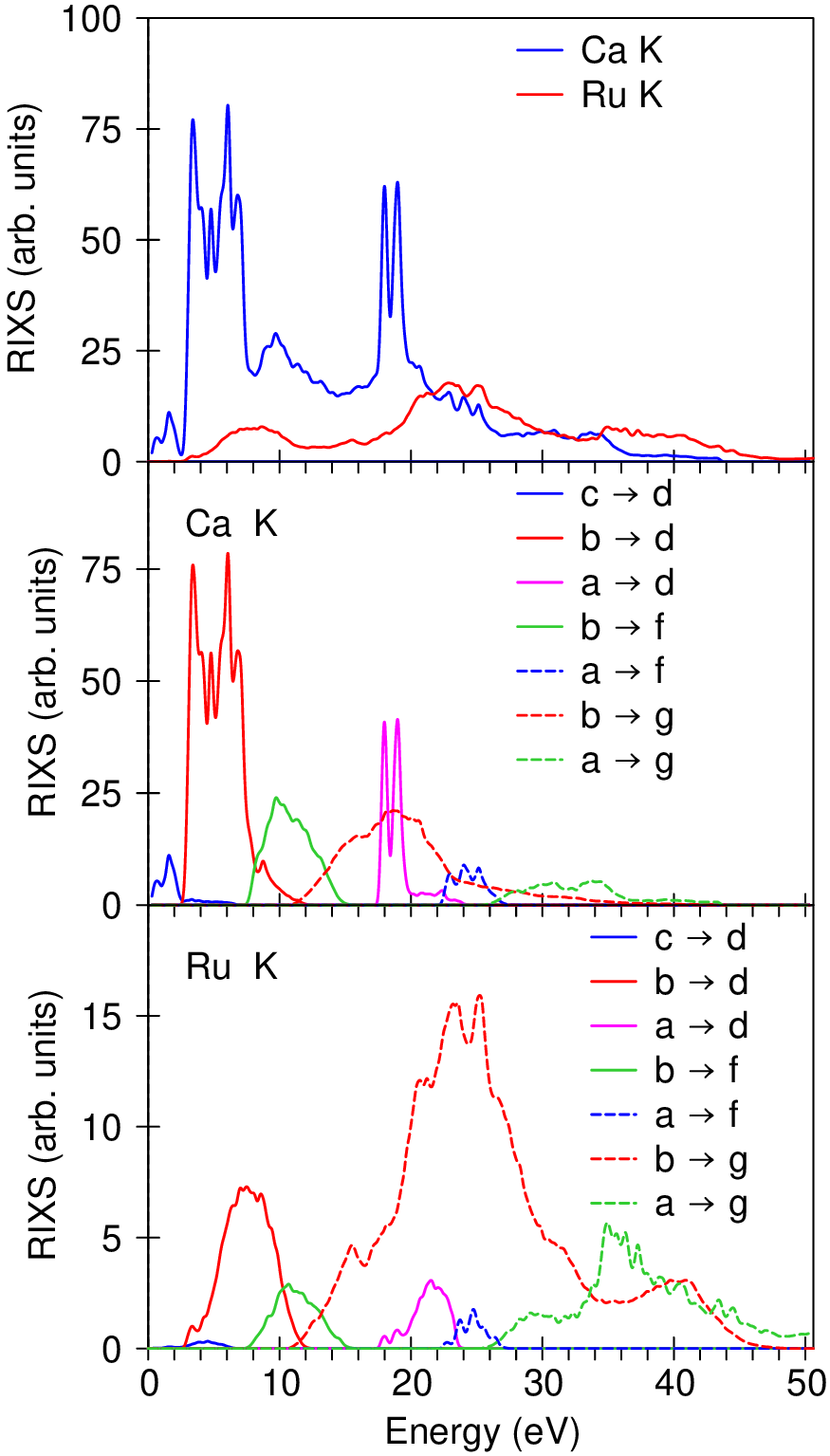}
\end{center}
\caption{\label{rixs_Ca_K_CRO}(Color online) The theoretically
  calculated RIXS spectra at the Ca and Ru $K$ edges (the upper panel)
  and partial contributions to the RIXS spectra from different
  interband transitions (the lower panel) in Ca$_3$Ru$_2$O$_7$
  calculated in the GGA+SO+$U$ approach ($U_{eff}$ = 1.3 eV). }
\end{figure}

Figure \ref{rixs_Ca_K_CRO} (the upper panel) presents the theoretically
calculated Ca and Ru $K$ RIXS spectra in Ca$_3$Ru$_2$O$_7$. The spectra
significantly differ from each other. The partial contributions from
different interband transitions are also presented in Fig. \ref{rixs_Ca_K_CRO}
at the Ru (the lower panel) and Ca (the middle panel) $K$ edges. The $c$
$\rightarrow$ $d$ ({\tg} $\rightarrow$ {\tg}) transitions are visible at the
Ca site but they are very small at the Ru one. The transitions to empty {\tg} UEB
states from the $p$ states which are hybridized with corresponding oxygen 2$p$
and 2$s$ states ($b$ $\rightarrow$ $d$ and $a$ $\rightarrow$ $d$,
respectively) are very sharp and intense at the Ca site and suppressed at the
Ru one due to the very large width of the Ru 1$s$ core level. The largest
contribution to the Ru $K$ RIXS spectrum comes from $b$ $\rightarrow$ $g$
transitions. They are also shifted towards higher energy in comparison with
the Ca $K$ spectrum due to the corresponding shift of empty Ru 5$p$ states in
respect to Ca 4$p$ states (see Fig. \ref{PDOS Ca_Ru_p}).

Experimental measurements of the XAS, XMCD, and RIXS spectra at the Ca and
Ru $K$ edges are also highly desirable.

\section{Conclusions}

Our calculations show that Ca$_3$Ru$_2$O$_7$ has a novel Mott insulating
state, which is induced by the interplay of SO coupling and Coulomb repulsion.

The electronic and magnetic properties of Ca$_3$Ru$_2$O$_7$ oxide have
been investigated theoretically using first-principle calculations in
the frame of the fully relativistic spin-polarized Dirac LMTO
band-structure method in order to understand the importance of Coulomb
interaction and spin-orbit coupling. The GGA approximation produces a
metallic ground state in Ca$_3$Ru$_2$O$_7$ in contradiction with
experimental data which show that this oxide is a Mott-like AFM
insulator. To produce the correct ground state one has to take into
account strong Coulomb correlations. The GGA+SO+$U$ approach
shifts the occupied and empty {\tg} bands downward and upward,
respectively, by $U_{eff}$/2 producing a dielectric ground state in
Ca$_3$Ru$_2$O$_7$ with a direct energy gap of 0.215 eV and an indirect
one of 0.053 eV. The energy gap is increased with increasing Hubbard
$U$. We found that in the ground state Ca$_3$Ru$_2$O$_7$ possesses an
AFM-$b$ magnetic structure with the Ru spin moments ordered
antiferromgnetically along the $b$ axis in consistence with
experimental data. The Ru spin and orbital moments are parallel in
Ca$_3$Ru$_2$O$_7$ in accordance with Hund's third rule. The spin and
orbital parts of the magnetic moments for the Ca ions in
Ca$_3$Ru$_2$O$_7$ are found to be very small. The magnetic moments at
the oxygen sites are also relatively small with the largest spin moments
at the O$_4$ site and the smallest ones at the O$_2$ site.

We have investigated theoretically the XAS, XMCD, and RIXS spectra at
the Ru and Ca $L_3$, $M_3$, and $K$ edges as well as at the O $K$ edge
of the Ca$_3$Ru$_2$O$_7$ oxide. The calculated results are in good
agreement with available experimental data. We have found that the
best agreement between the theory and experiment in the RIXS spectra
at the Ru $L_3$ edge can be achieved for $U_{eff}$ = 1.3 eV. The
calculations with larger values of $U_{eff}$ shift the RIXS spectrum
towards higher energies.

The experimentally measured RIXS spectrum of Ca$_3$Ru$_2$O$_7$ at the
Ru $L_3$ edge possesses a sharp feature below 2 eV corresponding to
transitions within Ru {\tg} levels. The excitation located from 2 to 4
eV is due to {\tg} $\rightarrow$ {\eg} transitions. The third wide
structure situated at 4.5$-$11 eV appears due to transitions between
Ru 4$d_{\rm{O}}$ states and {\eg} and {\tg} states. The RIXS spectra
at the $L_3$ and $M_3$ edges have to possess similar shapes because
they reflect similar interband transitions. This is true indeed in the
case of the Ru $L_3$ and $M_3$ RIXS spectra. However, the shapes of
the $L_3$ and $M_3$ RIXS spectra at the Ca site quite differ from each
other. Such a puzzle can be explained by the difference in the widths
of corresponding core-levels. The significantly larger width of the
3$p_{3/2}$ core-level leads to the suppression of the low energy fine
structures in the Ca $M_3$ RIXS spectrum.

The RIXS spectrum at the O $K$ edge consists of three major inelastic
excitations. We found that the two low energy features $\le$2.5 eV are
due to interband transitions between occupied and empty O$_{\tg}$
states which appear due to the strong hybridization between oxygen
2$p$ and Ru {\tg} states in close vicinity to the Fermi level. The
next major peak at 2.5$-$8 eV reflects interband transitions between
occupied O 2$p$ and empty oxygen states which originate from the
hybridization with Ru {\tg} states. Rather weak O$_{2p}$ $\rightarrow$
O$_{e_g}$ transitions occupy a wide energy interval between 4 and 11
eV. Although the partial Ca 4$p$ and Ru 5$p$ densities of states
possess similar energy band structures the RIXS spectra at the Ca and
Ru $K$ edges significantly differ from each other due to the
difference in the widths of corresponding 1$s$ core levels.

\section*{Acknowledgments}

We are thankful to Dr. Alexander Yaresko from the Max Planck Institute FKF in
Stuttgart for helpful discussions.


\begin{thebibliography}{52}
\expandafter\ifx\csname natexlab\endcsname\relax\def\natexlab#1{#1}\fi
\expandafter\ifx\csname bibnamefont\endcsname\relax
  \def\bibnamefont#1{#1}\fi
\expandafter\ifx\csname bibfnamefont\endcsname\relax
  \def\bibfnamefont#1{#1}\fi
\expandafter\ifx\csname citenamefont\endcsname\relax
  \def\citenamefont#1{#1}\fi
\expandafter\ifx\csname url\endcsname\relax
  \def\url#1{\texttt{#1}}\fi
\expandafter\ifx\csname urlprefix\endcsname\relax\def\urlprefix{URL }\fi
\providecommand{\bibinfo}[2]{#2}
\providecommand{\eprint}[2][]{\url{#2}}

\bibitem[{\citenamefont{Maeno et~al.}(1994)\citenamefont{Maeno, Hashimoto,
  Yoshida, Nishizaki, Fujita, Bednorz, and Lichtenberg}}]{MHY+94}
\bibinfo{author}{\bibfnamefont{Y.}~\bibnamefont{Maeno}},
  \bibinfo{author}{\bibfnamefont{H.}~\bibnamefont{Hashimoto}},
  \bibinfo{author}{\bibfnamefont{K.}~\bibnamefont{Yoshida}},
  \bibinfo{author}{\bibfnamefont{S.}~\bibnamefont{Nishizaki}},
  \bibinfo{author}{\bibfnamefont{T.}~\bibnamefont{Fujita}},
  \bibinfo{author}{\bibfnamefont{J.~G.} \bibnamefont{Bednorz}},
  \bibnamefont{and}
  \bibinfo{author}{\bibfnamefont{F.}~\bibnamefont{Lichtenberg}},
  \bibinfo{journal}{Nature (London)} \textbf{\bibinfo{volume}{372}},
  \bibinfo{pages}{532} (\bibinfo{year}{1994}).

\bibitem[{\citenamefont{Maeno et~al.}(2001)\citenamefont{Maeno, Rice, and
  Sigrist}}]{MRS01}
\bibinfo{author}{\bibfnamefont{Y.}~\bibnamefont{Maeno}},
  \bibinfo{author}{\bibfnamefont{T.~M.} \bibnamefont{Rice}}, \bibnamefont{and}
  \bibinfo{author}{\bibfnamefont{M.}~\bibnamefont{Sigrist}},
  \bibinfo{journal}{Phys. Today} \textbf{\bibinfo{volume}{54}},
  \bibinfo{pages}{42} (\bibinfo{year}{2001}).

\bibitem[{\citenamefont{Cao et~al.}(1997{\natexlab{a}})\citenamefont{Cao,
  McCall, Shepard, Crow, and Guertin}}]{CMS+97}
\bibinfo{author}{\bibfnamefont{G.}~\bibnamefont{Cao}},
  \bibinfo{author}{\bibfnamefont{S.}~\bibnamefont{McCall}},
  \bibinfo{author}{\bibfnamefont{M.}~\bibnamefont{Shepard}},
  \bibinfo{author}{\bibfnamefont{J.~E.} \bibnamefont{Crow}}, \bibnamefont{and}
  \bibinfo{author}{\bibfnamefont{R.~P.} \bibnamefont{Guertin}},
  \bibinfo{journal}{Phys. Rev. B} \textbf{\bibinfo{volume}{56}},
  \bibinfo{pages}{321} (\bibinfo{year}{1997}{\natexlab{a}}).

\bibitem[{\citenamefont{Klein et~al.}(1999)\citenamefont{Klein, Antognazza,
  Geballe, Beasley, and Kapitulnik}}]{KAG+99}
\bibinfo{author}{\bibfnamefont{L.}~\bibnamefont{Klein}},
  \bibinfo{author}{\bibfnamefont{L.}~\bibnamefont{Antognazza}},
  \bibinfo{author}{\bibfnamefont{T.~H.} \bibnamefont{Geballe}},
  \bibinfo{author}{\bibfnamefont{M.~R.} \bibnamefont{Beasley}},
  \bibnamefont{and}
  \bibinfo{author}{\bibfnamefont{A.}~\bibnamefont{Kapitulnik}},
  \bibinfo{journal}{Phys. Rev. B} \textbf{\bibinfo{volume}{60}},
  \bibinfo{pages}{1448} (\bibinfo{year}{1999}).

\bibitem[{\citenamefont{Shaked et~al.}(2000)\citenamefont{Shaked, Jorgensen,
  Chmaissem, Ikeda, and Maeno}}]{SJC+00}
\bibinfo{author}{\bibfnamefont{H.}~\bibnamefont{Shaked}},
  \bibinfo{author}{\bibfnamefont{J.~D.} \bibnamefont{Jorgensen}},
  \bibinfo{author}{\bibfnamefont{O.}~\bibnamefont{Chmaissem}},
  \bibinfo{author}{\bibfnamefont{S.}~\bibnamefont{Ikeda}}, \bibnamefont{and}
  \bibinfo{author}{\bibfnamefont{Y.}~\bibnamefont{Maeno}}, \bibinfo{journal}{J.
  Solid State Chem.} \textbf{\bibinfo{volume}{154}}, \bibinfo{pages}{361}
  (\bibinfo{year}{2000}).

\bibitem[{\citenamefont{Cao et~al.}(2000)\citenamefont{Cao, Abbound, McCall,
  Crow, and Guertin}}]{CAM+00}
\bibinfo{author}{\bibfnamefont{G.}~\bibnamefont{Cao}},
  \bibinfo{author}{\bibfnamefont{K.}~\bibnamefont{Abbound}},
  \bibinfo{author}{\bibfnamefont{S.}~\bibnamefont{McCall}},
  \bibinfo{author}{\bibfnamefont{J.~E.} \bibnamefont{Crow}}, \bibnamefont{and}
  \bibinfo{author}{\bibfnamefont{R.~P.} \bibnamefont{Guertin}},
  \bibinfo{journal}{Phys. Rev. B} \textbf{\bibinfo{volume}{62}},
  \bibinfo{pages}{998} (\bibinfo{year}{2000}).

\bibitem[{\citenamefont{Yoshida et~al.}(2005)\citenamefont{Yoshida, Ikeda,
  Matsuhata, Shirakawa, Lee, and Katano}}]{YIM+05}
\bibinfo{author}{\bibfnamefont{Y.}~\bibnamefont{Yoshida}},
  \bibinfo{author}{\bibfnamefont{S.~I.} \bibnamefont{Ikeda}},
  \bibinfo{author}{\bibfnamefont{H.}~\bibnamefont{Matsuhata}},
  \bibinfo{author}{\bibfnamefont{N.}~\bibnamefont{Shirakawa}},
  \bibinfo{author}{\bibfnamefont{C.~H.} \bibnamefont{Lee}}, \bibnamefont{and}
  \bibinfo{author}{\bibfnamefont{S.}~\bibnamefont{Katano}},
  \bibinfo{journal}{Phys. Rev. B} \textbf{\bibinfo{volume}{72}},
  \bibinfo{pages}{054412} (\bibinfo{year}{2005}).

\bibitem[{\citenamefont{Borzi et~al.}(2008)\citenamefont{Borzi, Grigera,
  Farrell, Perry, Lister, Lee, Tennant, Maeno, and Mackenzie}}]{BGF+06}
\bibinfo{author}{\bibfnamefont{R.~A.} \bibnamefont{Borzi}},
  \bibinfo{author}{\bibfnamefont{S.~A.} \bibnamefont{Grigera}},
  \bibinfo{author}{\bibfnamefont{J.}~\bibnamefont{Farrell}},
  \bibinfo{author}{\bibfnamefont{R.~S.} \bibnamefont{Perry}},
  \bibinfo{author}{\bibfnamefont{S.~J.~S.} \bibnamefont{Lister}},
  \bibinfo{author}{\bibfnamefont{S.~L.} \bibnamefont{Lee}},
  \bibinfo{author}{\bibfnamefont{D.~A.} \bibnamefont{Tennant}},
  \bibinfo{author}{\bibfnamefont{Y.}~\bibnamefont{Maeno}}, \bibnamefont{and}
  \bibinfo{author}{\bibfnamefont{A.~P.} \bibnamefont{Mackenzie}},
  \bibinfo{journal}{Science} \textbf{\bibinfo{volume}{315}},
  \bibinfo{pages}{214} (\bibinfo{year}{2008}).

\bibitem[{\citenamefont{Ikeda et~al.}(2000)\citenamefont{Ikeda, Maeno,
  Nakatsuji, Kosaka, and Uwatoko}}]{IMN+00}
\bibinfo{author}{\bibfnamefont{S.~I.} \bibnamefont{Ikeda}},
  \bibinfo{author}{\bibfnamefont{Y.}~\bibnamefont{Maeno}},
  \bibinfo{author}{\bibfnamefont{S.}~\bibnamefont{Nakatsuji}},
  \bibinfo{author}{\bibfnamefont{M.}~\bibnamefont{Kosaka}}, \bibnamefont{and}
  \bibinfo{author}{\bibfnamefont{Y.}~\bibnamefont{Uwatoko}},
  \bibinfo{journal}{Phys. Rev. B} \textbf{\bibinfo{volume}{62}},
  \bibinfo{pages}{6089R} (\bibinfo{year}{2000}).

\bibitem[{\citenamefont{Cao et~al.}(1997{\natexlab{b}})\citenamefont{Cao,
  McCall, Crow, and Guertin}}]{CMC+97}
\bibinfo{author}{\bibfnamefont{G.}~\bibnamefont{Cao}},
  \bibinfo{author}{\bibfnamefont{S.}~\bibnamefont{McCall}},
  \bibinfo{author}{\bibfnamefont{J.~E.} \bibnamefont{Crow}}, \bibnamefont{and}
  \bibinfo{author}{\bibfnamefont{R.~P.} \bibnamefont{Guertin}},
  \bibinfo{journal}{Phys. Rev. Lett.} \textbf{\bibinfo{volume}{78}},
  \bibinfo{pages}{17151} (\bibinfo{year}{1997}{\natexlab{b}}).

\bibitem[{\citenamefont{Ke et~al.}(2011)\citenamefont{Ke, Hong, Peng, Nagler,
  Granroth, Lumsden, and Mao}}]{KHP+11}
\bibinfo{author}{\bibfnamefont{X.}~\bibnamefont{Ke}},
  \bibinfo{author}{\bibfnamefont{T.}~\bibnamefont{Hong}},
  \bibinfo{author}{\bibfnamefont{J.}~\bibnamefont{Peng}},
  \bibinfo{author}{\bibfnamefont{S.~E.} \bibnamefont{Nagler}},
  \bibinfo{author}{\bibfnamefont{G.~E.} \bibnamefont{Granroth}},
  \bibinfo{author}{\bibfnamefont{M.~D.} \bibnamefont{Lumsden}},
  \bibnamefont{and} \bibinfo{author}{\bibfnamefont{Z.~Q.} \bibnamefont{Mao}},
  \bibinfo{journal}{Phys. Rev. B} \textbf{\bibinfo{volume}{84}},
  \bibinfo{pages}{014422} (\bibinfo{year}{2011}).

\bibitem[{\citenamefont{Ohmichi et~al.}(2004)\citenamefont{Ohmichi, Yoshida,
  Ikeda, Shirakawa, and Osada}}]{OYI+04}
\bibinfo{author}{\bibfnamefont{E.}~\bibnamefont{Ohmichi}},
  \bibinfo{author}{\bibfnamefont{Y.}~\bibnamefont{Yoshida}},
  \bibinfo{author}{\bibfnamefont{S.~I.} \bibnamefont{Ikeda}},
  \bibinfo{author}{\bibfnamefont{N.}~\bibnamefont{Shirakawa}},
  \bibnamefont{and} \bibinfo{author}{\bibfnamefont{T.}~\bibnamefont{Osada}},
  \bibinfo{journal}{Phys. Rev. B} \textbf{\bibinfo{volume}{70}},
  \bibinfo{pages}{104414} (\bibinfo{year}{2004}).

\bibitem[{\citenamefont{Yoshida et~al.}(2004)\citenamefont{Yoshida, Nagai,
  Ikeda, Shirakawa, Kosaka, and Mori}}]{YNI+04}
\bibinfo{author}{\bibfnamefont{Y.}~\bibnamefont{Yoshida}},
  \bibinfo{author}{\bibfnamefont{I.}~\bibnamefont{Nagai}},
  \bibinfo{author}{\bibfnamefont{S.~I.} \bibnamefont{Ikeda}},
  \bibinfo{author}{\bibfnamefont{N.}~\bibnamefont{Shirakawa}},
  \bibinfo{author}{\bibfnamefont{M.}~\bibnamefont{Kosaka}}, \bibnamefont{and}
  \bibinfo{author}{\bibfnamefont{N.}~\bibnamefont{Mori}},
  \bibinfo{journal}{Phys. Rev. B} \textbf{\bibinfo{volume}{69}},
  \bibinfo{pages}{220411} (\bibinfo{year}{2004}).

\bibitem[{\citenamefont{Dashwood et~al.}(2020)\citenamefont{Dashwood, Veiga,
  Faure, Vale, Porter, Collins, Manuel, Khalyavin, Orlandi, Perry
  et~al.}}]{DVF+20}
\bibinfo{author}{\bibfnamefont{C.~D.} \bibnamefont{Dashwood}},
  \bibinfo{author}{\bibfnamefont{L.~S.~I.} \bibnamefont{Veiga}},
  \bibinfo{author}{\bibfnamefont{Q.}~\bibnamefont{Faure}},
  \bibinfo{author}{\bibfnamefont{J.~G.} \bibnamefont{Vale}},
  \bibinfo{author}{\bibfnamefont{D.~G.} \bibnamefont{Porter}},
  \bibinfo{author}{\bibfnamefont{S.~P.} \bibnamefont{Collins}},
  \bibinfo{author}{\bibfnamefont{P.}~\bibnamefont{Manuel}},
  \bibinfo{author}{\bibfnamefont{D.~D.} \bibnamefont{Khalyavin}},
  \bibinfo{author}{\bibfnamefont{F.}~\bibnamefont{Orlandi}},
  \bibinfo{author}{\bibfnamefont{R.~S.} \bibnamefont{Perry}},
  \bibnamefont{et~al.}, \bibinfo{journal}{Phys. Rev. B}
  \textbf{\bibinfo{volume}{102}}, \bibinfo{pages}{180410(R)}
  (\bibinfo{year}{2020}).

\bibitem[{\citenamefont{Bao et~al.}(2008)\citenamefont{Bao, Mao, Qu, and
  Lynn}}]{BMQ+08}
\bibinfo{author}{\bibfnamefont{W.}~\bibnamefont{Bao}},
  \bibinfo{author}{\bibfnamefont{Z.~Q.} \bibnamefont{Mao}},
  \bibinfo{author}{\bibfnamefont{Z.}~\bibnamefont{Qu}}, \bibnamefont{and}
  \bibinfo{author}{\bibfnamefont{J.~W.} \bibnamefont{Lynn}},
  \bibinfo{journal}{Phys. Rev. Lett.} \textbf{\bibinfo{volume}{100}},
  \bibinfo{pages}{247203} (\bibinfo{year}{2008}).

\bibitem[{\citenamefont{Sokolov et~al.}(2019)\citenamefont{Sokolov, Kikugawa,
  Helm, Borrmann, Burkhardt, Cubitt, White, Ressouche, Bleuel, Kummer
  et~al.}}]{SKH+19}
\bibinfo{author}{\bibfnamefont{D.~A.} \bibnamefont{Sokolov}},
  \bibinfo{author}{\bibfnamefont{N.}~\bibnamefont{Kikugawa}},
  \bibinfo{author}{\bibfnamefont{T.}~\bibnamefont{Helm}},
  \bibinfo{author}{\bibfnamefont{H.}~\bibnamefont{Borrmann}},
  \bibinfo{author}{\bibfnamefont{U.}~\bibnamefont{Burkhardt}},
  \bibinfo{author}{\bibfnamefont{R.}~\bibnamefont{Cubitt}},
  \bibinfo{author}{\bibfnamefont{J.~S.} \bibnamefont{White}},
  \bibinfo{author}{\bibfnamefont{E.}~\bibnamefont{Ressouche}},
  \bibinfo{author}{\bibfnamefont{M.}~\bibnamefont{Bleuel}},
  \bibinfo{author}{\bibfnamefont{K.}~\bibnamefont{Kummer}},
  \bibnamefont{et~al.}, \bibinfo{journal}{Nat. Phys.}
  \textbf{\bibinfo{volume}{15}}, \bibinfo{pages}{671} (\bibinfo{year}{2019}).

\bibitem[{\citenamefont{McCall et~al.}(1998)\citenamefont{McCall, Cao, Crow,
  Harrison, Mielke, Lacerda, and Guertin}}]{MCC+98}
\bibinfo{author}{\bibfnamefont{S.}~\bibnamefont{McCall}},
  \bibinfo{author}{\bibfnamefont{G.}~\bibnamefont{Cao}},
  \bibinfo{author}{\bibfnamefont{J.~E.} \bibnamefont{Crow}},
  \bibinfo{author}{\bibfnamefont{N.}~\bibnamefont{Harrison}},
  \bibinfo{author}{\bibfnamefont{C.~H.} \bibnamefont{Mielke}},
  \bibinfo{author}{\bibfnamefont{A.~H.} \bibnamefont{Lacerda}},
  \bibnamefont{and} \bibinfo{author}{\bibfnamefont{R.~P.}
  \bibnamefont{Guertin}}, \bibinfo{journal}{Physica B}
  \textbf{\bibinfo{volume}{246}} (\bibinfo{year}{1998}).

\bibitem[{\citenamefont{Cao et~al.}(2003)\citenamefont{Cao, Balicas, Xin,
  Dagotto, Crow, Nelson, and Agterberg}}]{CBX+03}
\bibinfo{author}{\bibfnamefont{G.}~\bibnamefont{Cao}},
  \bibinfo{author}{\bibfnamefont{L.}~\bibnamefont{Balicas}},
  \bibinfo{author}{\bibfnamefont{Y.}~\bibnamefont{Xin}},
  \bibinfo{author}{\bibfnamefont{E.}~\bibnamefont{Dagotto}},
  \bibinfo{author}{\bibfnamefont{J.~E.} \bibnamefont{Crow}},
  \bibinfo{author}{\bibfnamefont{C.~S.} \bibnamefont{Nelson}},
  \bibnamefont{and} \bibinfo{author}{\bibfnamefont{D.~F.}
  \bibnamefont{Agterberg}}, \bibinfo{journal}{Phys. Rev. B}
  \textbf{\bibinfo{volume}{67}}, \bibinfo{pages}{060406(R)}
  (\bibinfo{year}{2003}).

\bibitem[{\citenamefont{Cao et~al.}(2004)\citenamefont{Cao, Lin, Balicas,
  Chikara, Crow, and Schlottmann}}]{CLB+04}
\bibinfo{author}{\bibfnamefont{G.}~\bibnamefont{Cao}},
  \bibinfo{author}{\bibfnamefont{X.~N.} \bibnamefont{Lin}},
  \bibinfo{author}{\bibfnamefont{L.}~\bibnamefont{Balicas}},
  \bibinfo{author}{\bibfnamefont{S.}~\bibnamefont{Chikara}},
  \bibinfo{author}{\bibfnamefont{J.~E.} \bibnamefont{Crow}}, \bibnamefont{and}
  \bibinfo{author}{\bibfnamefont{P.}~\bibnamefont{Schlottmann}},
  \bibinfo{journal}{New J. Phys.} \textbf{\bibinfo{volume}{6}},
  \bibinfo{pages}{159} (\bibinfo{year}{2004}).

\bibitem[{\citenamefont{Puchkov et~al.}(1998)\citenamefont{Puchkov, Schabel,
  Basov, Startseva, Cao, Timusk, and Shen}}]{PSB+98}
\bibinfo{author}{\bibfnamefont{A.~V.} \bibnamefont{Puchkov}},
  \bibinfo{author}{\bibfnamefont{M.~C.} \bibnamefont{Schabel}},
  \bibinfo{author}{\bibfnamefont{D.~N.} \bibnamefont{Basov}},
  \bibinfo{author}{\bibfnamefont{T.}~\bibnamefont{Startseva}},
  \bibinfo{author}{\bibfnamefont{G.}~\bibnamefont{Cao}},
  \bibinfo{author}{\bibfnamefont{T.}~\bibnamefont{Timusk}}, \bibnamefont{and}
  \bibinfo{author}{\bibfnamefont{Z.-X.} \bibnamefont{Shen}},
  \bibinfo{journal}{Phys. Rev. Lett.} \textbf{\bibinfo{volume}{81}},
  \bibinfo{pages}{144} (\bibinfo{year}{1998}).

\bibitem[{\citenamefont{Cao et~al.}(1997{\natexlab{c}})\citenamefont{Cao,
  McCall, Crow, and Guertin}}]{CMC+99}
\bibinfo{author}{\bibfnamefont{G.}~\bibnamefont{Cao}},
  \bibinfo{author}{\bibfnamefont{S.~C.} \bibnamefont{McCall}},
  \bibinfo{author}{\bibfnamefont{J.~E.} \bibnamefont{Crow}}, \bibnamefont{and}
  \bibinfo{author}{\bibfnamefont{R.~P.} \bibnamefont{Guertin}},
  \bibinfo{journal}{Phys. Rev. B} \textbf{\bibinfo{volume}{56}},
  \bibinfo{pages}{5387} (\bibinfo{year}{1997}{\natexlab{c}}).

\bibitem[{\citenamefont{Lee et~al.}(2007)\citenamefont{Lee, Moon, Yang, Yu,
  Schade, Yoshida, Ikeda, and Noh}}]{LMY+07}
\bibinfo{author}{\bibfnamefont{J.~S.} \bibnamefont{Lee}},
  \bibinfo{author}{\bibfnamefont{S.~J.} \bibnamefont{Moon}},
  \bibinfo{author}{\bibfnamefont{B.~J.} \bibnamefont{Yang}},
  \bibinfo{author}{\bibfnamefont{J.}~\bibnamefont{Yu}},
  \bibinfo{author}{\bibfnamefont{U.}~\bibnamefont{Schade}},
  \bibinfo{author}{\bibfnamefont{Y.}~\bibnamefont{Yoshida}},
  \bibinfo{author}{\bibfnamefont{S.-I.} \bibnamefont{Ikeda}}, \bibnamefont{and}
  \bibinfo{author}{\bibfnamefont{T.~W.} \bibnamefont{Noh}},
  \bibinfo{journal}{Phys. Rev. Lett.} \textbf{\bibinfo{volume}{98}},
  \bibinfo{pages}{097403} (\bibinfo{year}{2007}).

\bibitem[{\citenamefont{Liu et~al.}(1999)\citenamefont{Liu, Yoon, Cooper, Cao,
  and Crow}}]{LYC+99}
\bibinfo{author}{\bibfnamefont{H.~L.} \bibnamefont{Liu}},
  \bibinfo{author}{\bibfnamefont{S.}~\bibnamefont{Yoon}},
  \bibinfo{author}{\bibfnamefont{S.~L.} \bibnamefont{Cooper}},
  \bibinfo{author}{\bibfnamefont{G.}~\bibnamefont{Cao}}, \bibnamefont{and}
  \bibinfo{author}{\bibfnamefont{J.~E.} \bibnamefont{Crow}},
  \bibinfo{journal}{Phys. Rev. B} \textbf{\bibinfo{volume}{60}},
  \bibinfo{pages}{6980(R)} (\bibinfo{year}{1999}).

\bibitem[{\citenamefont{Mazin and Singh}(1997)}]{MaSi97}
\bibinfo{author}{\bibfnamefont{I.~I.} \bibnamefont{Mazin}} \bibnamefont{and}
  \bibinfo{author}{\bibfnamefont{D.~J.} \bibnamefont{Singh}},
  \bibinfo{journal}{Phys. Rev. B} \textbf{\bibinfo{volume}{56}},
  \bibinfo{pages}{2556} (\bibinfo{year}{1997}).

\bibitem[{\citenamefont{Mizokawa et~al.}(2001)\citenamefont{Mizokawa, Tjeng,
  Sawatzky, Ghiringhelli, Tjernberg, Brookes, Fukazawa, Nakatsuji, and
  Maeno}}]{MTS+01}
\bibinfo{author}{\bibfnamefont{T.}~\bibnamefont{Mizokawa}},
  \bibinfo{author}{\bibfnamefont{L.~H.} \bibnamefont{Tjeng}},
  \bibinfo{author}{\bibfnamefont{G.~A.} \bibnamefont{Sawatzky}},
  \bibinfo{author}{\bibfnamefont{G.}~\bibnamefont{Ghiringhelli}},
  \bibinfo{author}{\bibfnamefont{O.}~\bibnamefont{Tjernberg}},
  \bibinfo{author}{\bibfnamefont{N.~B.} \bibnamefont{Brookes}},
  \bibinfo{author}{\bibfnamefont{H.}~\bibnamefont{Fukazawa}},
  \bibinfo{author}{\bibfnamefont{S.}~\bibnamefont{Nakatsuji}},
  \bibnamefont{and} \bibinfo{author}{\bibfnamefont{Y.}~\bibnamefont{Maeno}},
  \bibinfo{journal}{Phys. Rev. Lett.} \textbf{\bibinfo{volume}{87}},
  \bibinfo{pages}{077202} (\bibinfo{year}{2001}).

\bibitem[{\citenamefont{Fang et~al.}(2004)\citenamefont{Fang, Nagaosa, and
  Terakura}}]{FNT04}
\bibinfo{author}{\bibfnamefont{Z.}~\bibnamefont{Fang}},
  \bibinfo{author}{\bibfnamefont{N.}~\bibnamefont{Nagaosa}}, \bibnamefont{and}
  \bibinfo{author}{\bibfnamefont{K.}~\bibnamefont{Terakura}},
  \bibinfo{journal}{Phys. Rev. B} \textbf{\bibinfo{volume}{69}},
  \bibinfo{pages}{045116} (\bibinfo{year}{2004}).

\bibitem[{\citenamefont{Liu et~al.}(2008)\citenamefont{Liu, Antonov, Jepsen,
  and Andersen}}]{LAJ+08}
\bibinfo{author}{\bibfnamefont{G.-Q.} \bibnamefont{Liu}},
  \bibinfo{author}{\bibfnamefont{V.~N.} \bibnamefont{Antonov}},
  \bibinfo{author}{\bibfnamefont{O.}~\bibnamefont{Jepsen}}, \bibnamefont{and}
  \bibinfo{author}{\bibfnamefont{O.~K.} \bibnamefont{Andersen}},
  \bibinfo{journal}{Phys. Rev. Lett.} \textbf{\bibinfo{volume}{101}},
  \bibinfo{pages}{026408} (\bibinfo{year}{2008}).

\bibitem[{\citenamefont{Haverkort et~al.}(2008)\citenamefont{Haverkort,
  Elfimov, Tjeng, Sawatzky, and Damascelli}}]{HET+08}
\bibinfo{author}{\bibfnamefont{M.~W.} \bibnamefont{Haverkort}},
  \bibinfo{author}{\bibfnamefont{I.~S.} \bibnamefont{Elfimov}},
  \bibinfo{author}{\bibfnamefont{L.~H.} \bibnamefont{Tjeng}},
  \bibinfo{author}{\bibfnamefont{G.~A.} \bibnamefont{Sawatzky}},
  \bibnamefont{and}
  \bibinfo{author}{\bibfnamefont{A.}~\bibnamefont{Damascelli}},
  \bibinfo{journal}{Phys. Rev. Lett.} \textbf{\bibinfo{volume}{101}},
  \bibinfo{pages}{026406} (\bibinfo{year}{2008}).

\bibitem[{\citenamefont{Bertinshaw et~al.}(2021)\citenamefont{Bertinshaw,
  Krautloher, Suzuki, Takahashi, Ivanov, Yava, Kim, Gretarsson, and
  Keimer}}]{BKS+21}
\bibinfo{author}{\bibfnamefont{J.}~\bibnamefont{Bertinshaw}},
  \bibinfo{author}{\bibfnamefont{M.}~\bibnamefont{Krautloher}},
  \bibinfo{author}{\bibfnamefont{H.}~\bibnamefont{Suzuki}},
  \bibinfo{author}{\bibfnamefont{H.}~\bibnamefont{Takahashi}},
  \bibinfo{author}{\bibfnamefont{A.}~\bibnamefont{Ivanov}},
  \bibinfo{author}{\bibfnamefont{H.}~\bibnamefont{Yava}},
  \bibinfo{author}{\bibfnamefont{B.~J.} \bibnamefont{Kim}},
  \bibinfo{author}{\bibfnamefont{H.}~\bibnamefont{Gretarsson}},
  \bibnamefont{and} \bibinfo{author}{\bibfnamefont{B.}~\bibnamefont{Keimer}},
  \bibinfo{journal}{Phys. Rev. B} \textbf{\bibinfo{volume}{103}},
  \bibinfo{pages}{085108} (\bibinfo{year}{2021}).

\bibitem[{\citenamefont{Singh and Auluck}(2006)}]{SiAu06}
\bibinfo{author}{\bibfnamefont{D.~J.} \bibnamefont{Singh}} \bibnamefont{and}
  \bibinfo{author}{\bibfnamefont{S.}~\bibnamefont{Auluck}},
  \bibinfo{journal}{Phys. Rev. Lett.} \textbf{\bibinfo{volume}{96}},
  \bibinfo{pages}{097203} (\bibinfo{year}{2006}).

\bibitem[{\citenamefont{Liu}(2011)}]{Liu11}
\bibinfo{author}{\bibfnamefont{G.-Q.} \bibnamefont{Liu}},
  \bibinfo{journal}{Phys. Rev. B} \textbf{\bibinfo{volume}{84}},
  \bibinfo{pages}{235137} (\bibinfo{year}{2011}).

\bibitem[{\citenamefont{Jin and Ku}(2018)}]{JiKu18}
\bibinfo{author}{\bibfnamefont{Z.}~\bibnamefont{Jin}} \bibnamefont{and}
  \bibinfo{author}{\bibfnamefont{W.}~\bibnamefont{Ku}},
  \bibinfo{journal}{preprint arXiv:1809.04589}  (\bibinfo{year}{2018}).

\bibitem[{\citenamefont{Markovi et~al.}(2018)\citenamefont{Markovi, Watson,
  Clark, Mazzola, Morales, Hooley, Rosner, Polley, Balasubramanian, Mukherjee
  et~al.}}]{MWC+20}
\bibinfo{author}{\bibfnamefont{I.}~\bibnamefont{Markovi}},
  \bibinfo{author}{\bibfnamefont{M.~D.} \bibnamefont{Watson}},
  \bibinfo{author}{\bibfnamefont{O.~J.} \bibnamefont{Clark}},
  \bibinfo{author}{\bibfnamefont{F.}~\bibnamefont{Mazzola}},
  \bibinfo{author}{\bibfnamefont{E.~A.} \bibnamefont{Morales}},
  \bibinfo{author}{\bibfnamefont{C.~A.} \bibnamefont{Hooley}},
  \bibinfo{author}{\bibfnamefont{H.}~\bibnamefont{Rosner}},
  \bibinfo{author}{\bibfnamefont{C.~M.} \bibnamefont{Polley}},
  \bibinfo{author}{\bibfnamefont{T.}~\bibnamefont{Balasubramanian}},
  \bibinfo{author}{\bibfnamefont{S.}~\bibnamefont{Mukherjee}},
  \bibnamefont{et~al.}, \bibinfo{journal}{preprint arXiv:2001.09499}
  (\bibinfo{year}{2018}).

\bibitem[{\citenamefont{Puggioni et~al.}(2020)\citenamefont{Puggioni, Horio,
  Chang, and Rondinelli}}]{PHC+20}
\bibinfo{author}{\bibfnamefont{D.}~\bibnamefont{Puggioni}},
  \bibinfo{author}{\bibfnamefont{M.}~\bibnamefont{Horio}},
  \bibinfo{author}{\bibfnamefont{J.}~\bibnamefont{Chang}}, \bibnamefont{and}
  \bibinfo{author}{\bibfnamefont{J.~M.} \bibnamefont{Rondinelli}},
  \bibinfo{journal}{Phys. Rev. Reseach} \textbf{\bibinfo{volume}{2}},
  \bibinfo{pages}{023141} (\bibinfo{year}{2020}).

\bibitem[{\citenamefont{von Arx et~al.}(2020)\citenamefont{von Arx, Forte,
  Horio, Granata, Wang, Das, Sassa, Fittipaldi, Fatuzzo, Ivashko
  et~al.}}]{AFH+20}
\bibinfo{author}{\bibfnamefont{K.}~\bibnamefont{von Arx}},
  \bibinfo{author}{\bibfnamefont{F.}~\bibnamefont{Forte}},
  \bibinfo{author}{\bibfnamefont{M.}~\bibnamefont{Horio}},
  \bibinfo{author}{\bibfnamefont{V.}~\bibnamefont{Granata}},
  \bibinfo{author}{\bibfnamefont{Q.}~\bibnamefont{Wang}},
  \bibinfo{author}{\bibfnamefont{L.}~\bibnamefont{Das}},
  \bibinfo{author}{\bibfnamefont{Y.}~\bibnamefont{Sassa}},
  \bibinfo{author}{\bibfnamefont{R.}~\bibnamefont{Fittipaldi}},
  \bibinfo{author}{\bibfnamefont{C.~G.} \bibnamefont{Fatuzzo}},
  \bibinfo{author}{\bibfnamefont{O.}~\bibnamefont{Ivashko}},
  \bibnamefont{et~al.}, \bibinfo{journal}{Phys. Rev. B}
  \textbf{\bibinfo{volume}{102}}, \bibinfo{pages}{235104}
  (\bibinfo{year}{2020}).

\bibitem[{\citenamefont{Ament et~al.}(2011)\citenamefont{Ament, van Veenendaal,
  Devereaux, Hill, and van~den Brink}}]{AVD+11}
\bibinfo{author}{\bibfnamefont{L.~J.~P.} \bibnamefont{Ament}},
  \bibinfo{author}{\bibfnamefont{M.}~\bibnamefont{van Veenendaal}},
  \bibinfo{author}{\bibfnamefont{T.~P.} \bibnamefont{Devereaux}},
  \bibinfo{author}{\bibfnamefont{J.~P.} \bibnamefont{Hill}}, \bibnamefont{and}
  \bibinfo{author}{\bibfnamefont{J.}~\bibnamefont{van~den Brink}},
  \bibinfo{journal}{Rev. Mod. Phys.} \textbf{\bibinfo{volume}{83}},
  \bibinfo{pages}{705} (\bibinfo{year}{2011}).

\bibitem[{\citenamefont{Nemoshkalenko et~al.}(1983)\citenamefont{Nemoshkalenko,
  Krasovskii, Antonov, Antonov, Fleck, Wonn, and Ziesche}}]{NKA+83}
\bibinfo{author}{\bibfnamefont{V.~V.} \bibnamefont{Nemoshkalenko}},
  \bibinfo{author}{\bibfnamefont{A.~E.} \bibnamefont{Krasovskii}},
  \bibinfo{author}{\bibfnamefont{V.~N.} \bibnamefont{Antonov}},
  \bibinfo{author}{\bibfnamefont{V.~N.} \bibnamefont{Antonov}},
  \bibinfo{author}{\bibfnamefont{U.}~\bibnamefont{Fleck}},
  \bibinfo{author}{\bibfnamefont{H.}~\bibnamefont{Wonn}}, \bibnamefont{and}
  \bibinfo{author}{\bibfnamefont{P.}~\bibnamefont{Ziesche}},
  \bibinfo{journal}{Phys. status solidi B} \textbf{\bibinfo{volume}{120}},
  \bibinfo{pages}{283} (\bibinfo{year}{1983}).

\bibitem[{\citenamefont{Arola et~al.}(1997)\citenamefont{Arola, Strange, and
  Gyorffy}}]{ASG97}
\bibinfo{author}{\bibfnamefont{E.}~\bibnamefont{Arola}},
  \bibinfo{author}{\bibfnamefont{P.}~\bibnamefont{Strange}}, \bibnamefont{and}
  \bibinfo{author}{\bibfnamefont{B.~L.} \bibnamefont{Gyorffy}},
  \bibinfo{journal}{Phys. Rev. B} \textbf{\bibinfo{volume}{55}},
  \bibinfo{pages}{472} (\bibinfo{year}{1997}).

\bibitem[{\citenamefont{Antonov et~al.}(2022)\citenamefont{Antonov, Kukusta,
  and Bekenov}}]{AKB22a}
\bibinfo{author}{\bibfnamefont{V.~N.} \bibnamefont{Antonov}},
  \bibinfo{author}{\bibfnamefont{D.~A.} \bibnamefont{Kukusta}},
  \bibnamefont{and} \bibinfo{author}{\bibfnamefont{L.~V.}
  \bibnamefont{Bekenov}}, \bibinfo{journal}{Phys. Rev. B}
  \textbf{\bibinfo{volume}{105}}, \bibinfo{pages}{155144}
  (\bibinfo{year}{2022}).

\bibitem[{\citenamefont{Guo et~al.}(1994)\citenamefont{Guo, Ebert, Temmerman,
  and Durham}}]{GET+94}
\bibinfo{author}{\bibfnamefont{G.~Y.} \bibnamefont{Guo}},
  \bibinfo{author}{\bibfnamefont{H.}~\bibnamefont{Ebert}},
  \bibinfo{author}{\bibfnamefont{W.~M.} \bibnamefont{Temmerman}},
  \bibnamefont{and} \bibinfo{author}{\bibfnamefont{P.~J.}
  \bibnamefont{Durham}}, \bibinfo{journal}{Phys. Rev. B}
  \textbf{\bibinfo{volume}{50}}, \bibinfo{pages}{3861} (\bibinfo{year}{1994}).

\bibitem[{\citenamefont{Antonov et~al.}(2004)\citenamefont{Antonov, Harmon, and
  Yaresko}}]{book:AHY04}
\bibinfo{author}{\bibfnamefont{V.}~\bibnamefont{Antonov}},
  \bibinfo{author}{\bibfnamefont{B.}~\bibnamefont{Harmon}}, \bibnamefont{and}
  \bibinfo{author}{\bibfnamefont{A.}~\bibnamefont{Yaresko}},
  \emph{\bibinfo{title}{Electronic Structure and Magneto-Optical Properties of
  Solids}} (\bibinfo{publisher}{Kluwer}, \bibinfo{address}{Dordrecht},
  \bibinfo{year}{2004}).

\bibitem[{\citenamefont{Arola et~al.}(2004)\citenamefont{Arola, Horne, Strange,
  Winter, Szotek, and Temmerman}}]{AHS+04}
\bibinfo{author}{\bibfnamefont{E.}~\bibnamefont{Arola}},
  \bibinfo{author}{\bibfnamefont{M.}~\bibnamefont{Horne}},
  \bibinfo{author}{\bibfnamefont{P.}~\bibnamefont{Strange}},
  \bibinfo{author}{\bibfnamefont{H.}~\bibnamefont{Winter}},
  \bibinfo{author}{\bibfnamefont{Z.}~\bibnamefont{Szotek}}, \bibnamefont{and}
  \bibinfo{author}{\bibfnamefont{W.~M.} \bibnamefont{Temmerman}},
  \bibinfo{journal}{Phys. Rev. B} \textbf{\bibinfo{volume}{70}},
  \bibinfo{pages}{235127} (\bibinfo{year}{2004}).

\bibitem[{\citenamefont{Antonov et~al.}(2006)\citenamefont{Antonov, Jepsen,
  Yaresko, and Shpak}}]{AJY+06}
\bibinfo{author}{\bibfnamefont{V.~N.} \bibnamefont{Antonov}},
  \bibinfo{author}{\bibfnamefont{O.}~\bibnamefont{Jepsen}},
  \bibinfo{author}{\bibfnamefont{A.~N.} \bibnamefont{Yaresko}},
  \bibnamefont{and} \bibinfo{author}{\bibfnamefont{A.~P.} \bibnamefont{Shpak}},
  \bibinfo{journal}{J. Appl. Phys.} \textbf{\bibinfo{volume}{100}},
  \bibinfo{pages}{043711} (\bibinfo{year}{2006}).

\bibitem[{\citenamefont{Antonov et~al.}(2007)\citenamefont{Antonov, Harmon,
  Yaresko, and Shpak}}]{AHY+07b}
\bibinfo{author}{\bibfnamefont{V.~N.} \bibnamefont{Antonov}},
  \bibinfo{author}{\bibfnamefont{B.~N.} \bibnamefont{Harmon}},
  \bibinfo{author}{\bibfnamefont{A.~N.} \bibnamefont{Yaresko}},
  \bibnamefont{and} \bibinfo{author}{\bibfnamefont{A.~P.} \bibnamefont{Shpak}},
  \bibinfo{journal}{Phys. Rev. B} \textbf{\bibinfo{volume}{75}},
  \bibinfo{pages}{184422} (\bibinfo{year}{2007}).

\bibitem[{\citenamefont{Antonov et~al.}(2010)\citenamefont{Antonov, Yaresko,
  and Jepsen}}]{AYJ10}
\bibinfo{author}{\bibfnamefont{V.~N.} \bibnamefont{Antonov}},
  \bibinfo{author}{\bibfnamefont{A.~N.} \bibnamefont{Yaresko}},
  \bibnamefont{and} \bibinfo{author}{\bibfnamefont{O.}~\bibnamefont{Jepsen}},
  \bibinfo{journal}{Phys. Rev. B} \textbf{\bibinfo{volume}{81}},
  \bibinfo{pages}{075209} (\bibinfo{year}{2010}).

\bibitem[{\citenamefont{Andersen}(1975)}]{And75}
\bibinfo{author}{\bibfnamefont{O.~K.} \bibnamefont{Andersen}},
  \bibinfo{journal}{Phys. Rev. B} \textbf{\bibinfo{volume}{12}},
  \bibinfo{pages}{3060} (\bibinfo{year}{1975}).

\bibitem[{\citenamefont{Perdew et~al.}(1996)\citenamefont{Perdew, Burke, and
  Ernzerhof}}]{PBE96}
\bibinfo{author}{\bibfnamefont{J.~P.} \bibnamefont{Perdew}},
  \bibinfo{author}{\bibfnamefont{K.}~\bibnamefont{Burke}}, \bibnamefont{and}
  \bibinfo{author}{\bibfnamefont{M.}~\bibnamefont{Ernzerhof}},
  \bibinfo{journal}{Phys. Rev. Lett.} \textbf{\bibinfo{volume}{77}},
  \bibinfo{pages}{3865} (\bibinfo{year}{1996}).

\bibitem[{\citenamefont{Bl\"ochl et~al.}(1994)\citenamefont{Bl\"ochl, Jepsen,
  and Andersen}}]{BJA94}
\bibinfo{author}{\bibfnamefont{P.~E.} \bibnamefont{Bl\"ochl}},
  \bibinfo{author}{\bibfnamefont{O.}~\bibnamefont{Jepsen}}, \bibnamefont{and}
  \bibinfo{author}{\bibfnamefont{O.~K.} \bibnamefont{Andersen}},
  \bibinfo{journal}{Phys. Rev. B} \textbf{\bibinfo{volume}{49}},
  \bibinfo{pages}{16223} (\bibinfo{year}{1994}).

\bibitem[{\citenamefont{Yaresko et~al.}(2003)\citenamefont{Yaresko, Antonov,
  and Fulde}}]{YAF03}
\bibinfo{author}{\bibfnamefont{A.~N.} \bibnamefont{Yaresko}},
  \bibinfo{author}{\bibfnamefont{V.~N.} \bibnamefont{Antonov}},
  \bibnamefont{and} \bibinfo{author}{\bibfnamefont{P.}~\bibnamefont{Fulde}},
  \bibinfo{journal}{Phys. Rev. B} \textbf{\bibinfo{volume}{67}},
  \bibinfo{pages}{155103} (\bibinfo{year}{2003}).

\bibitem[{\citenamefont{Dederichs et~al.}(1984)\citenamefont{Dederichs,
  Bl\"ugel, Zeller, and Akai}}]{DBZ+84}
\bibinfo{author}{\bibfnamefont{P.~H.} \bibnamefont{Dederichs}},
  \bibinfo{author}{\bibfnamefont{S.}~\bibnamefont{Bl\"ugel}},
  \bibinfo{author}{\bibfnamefont{R.}~\bibnamefont{Zeller}}, \bibnamefont{and}
  \bibinfo{author}{\bibfnamefont{H.}~\bibnamefont{Akai}},
  \bibinfo{journal}{Phys. Rev. Lett.} \textbf{\bibinfo{volume}{53}},
  \bibinfo{pages}{2512} (\bibinfo{year}{1984}).

\bibitem[{\citenamefont{Pickett et~al.}(1998)\citenamefont{Pickett, Erwin, and
  Ethridge}}]{PEE98}
\bibinfo{author}{\bibfnamefont{W.~E.} \bibnamefont{Pickett}},
  \bibinfo{author}{\bibfnamefont{S.~C.} \bibnamefont{Erwin}}, \bibnamefont{and}
  \bibinfo{author}{\bibfnamefont{E.~C.} \bibnamefont{Ethridge}},
  \bibinfo{journal}{Phys. Rev. B} \textbf{\bibinfo{volume}{58}},
  \bibinfo{pages}{1201} (\bibinfo{year}{1998}).

\bibitem[{\citenamefont{Campbell and Parr}(2001)}]{CaPa01}
\bibinfo{author}{\bibfnamefont{J.~L.} \bibnamefont{Campbell}} \bibnamefont{and}
  \bibinfo{author}{\bibfnamefont{T.}~\bibnamefont{Parr}}, \bibinfo{journal}{At.
  Data Nucl. Data Tables} \textbf{\bibinfo{volume}{77}}, \bibinfo{pages}{1}
  (\bibinfo{year}{2001}).

\end{thebibliography}

\newcommand{\noopsort}[1]{} \newcommand{\printfirst}[2]{#1}
  \newcommand{\singleletter}[1]{#1} \newcommand{\switchargs}[2]{#2#1}

\end{document}